\documentclass[aps,twocolumn,prd,showpacs,nofootinbib]{revtex4}
\usepackage{natbib}
\usepackage{amsmath}
\usepackage{graphicx}
\usepackage{subfigure}
\usepackage{bm}
\usepackage{amssymb,array}
\usepackage{latexsym}
\usepackage{feynmf}

\usepackage{ulem}
\usepackage{color}

\newcommand{\lta}{\lesssim}
\newcommand{\gta}{\gtrsim}

\usepackage{graphicx}
\newcommand{\cev}[1]{\reflectbox{\ensuremath{\vec{\reflectbox{\ensuremath{#1}}}}}}

\def\spose#1{\hbox to 0pt{#1\hss}}
\def\lta{\mathrel{\spose{\lower 3pt\hbox{$\mathchar"218$}}
     \raise 2.0pt\hbox{$\mathchar"13C$}}}
\def\gta{\mathrel{\spose{\lower 3pt\hbox{$\mathchar"218$}}
     \raise 2.0pt\hbox{$\mathchar"13E$}}}

\newcommand{\de}[2]{\kern - #1 em \mathrm{d} #2}

\newcommand{\usys}{\mathrm{sys}}
\newcommand{\ubath}{\mathrm{bath}}
\newcommand{\uint}{\mathrm{int}}
\newcommand{\ueff}{\mathrm{eff}}
\newcommand{\uIA}{\mathrm{IA}}
\newcommand{\uold}{\mathrm{old}}
\newcommand{\unew}{\mathrm{new}}
\newcommand{\Det}{\mathrm{Det}}

\def\be{\begin{equation}}
\def\ee{\end{equation}}
\def\bea{\begin{eqnarray}}
\def\eea{\end{eqnarray}}

\begin{document}
\title{Lagrangian Formulation of Stochastic Inflation: A Recursive Approach}

\author{Laurence Perreault Levasseur${}^1$}
\affiliation{DAMTP, Center for Mathematical Sciences, University of Cambridge \\ Cambridge CB3 0WA, United Kingdom}

%
%
%
%

\date{\today}

\begin{abstract}

We present a new, recursive approach to stochastic inflation which is self-consistent and solves multiple problems which plagued a certain number of previous studies, in particular in realistic contexts where the background spacetime is taken to be dynamical, where there is more than one field present, especially with a mass hierarchy, or where the role played by back-reaction is suspected to be important. We first review the formalism of stochastic inflation as it is usually heuristically presented, that is, deriving the Langevin equations from the field equations of motion, and summarize previous results on the subject. We demonstrate where inconsistent approximations to the Langevin equations are commonly made, and show how these can be avoided. This set up shares many similarities with quantum Brownian motion and out-of-equilibrium statistical quantum dynamics. We hence review how path integral techniques can be applied to the stochastic inflationary context. We show this formalism to be consistent with the standard approach, develop a natural perturbative expansion, and use it to calculate the one-loop corrected Langevin equations.

\end{abstract}

\maketitle

\maketitle

\let\thefootnote\relax\footnote{${}^1$l.perreault-levasseur@damtp.cam.ac.uk}

\section{Introduction}
\label{sec:intro}

	Since the publication of the seminal paper by Guth \cite{Guth1981}, where the idea of inflation was first proposed, inflationary cosmology has become one of the cornerstones of modern cosmology. Indeed, not only did it propose an elegant solution to both the homogeneity and flatness problems, but being the first compelling cosmological model allowing predictions based on causal physics about the structure of the Universe on large scales, this theory has dramatically changed the face of cosmology. The predictive power of inflation \cite{Mukhanov1981} together with the observations of the Cosmic Microwave Background (CMB) anisotropies with the COBE and the WMAP \cite{Bennett:2012fp} experiments, and very recently the Planck satellite \cite{Ade:2013xsa}, have truly embodied the last step in this transformation. Furthermore, by opening new avenues connecting theoretical physics to experiment, it has allowed substantial progress in ways to approach fundamental high energy physics. In fact, through the prediction of the shape and amplitude of the spectrum of primordial fluctuations of the CMB, inflation has provided us with the theoretical tools to analyze what can be thought of as a direct window into the Planck scale. In this sense, CMB experiments give us a direct access to the quantum world at energy scales far beyond what is accessible in Earth-based high energy particle physics experiments.
	
	With the ever-increasing precision of the experiments probing this window into the very beginning of the Universe, it becomes more and more critical to develop solid and self-consistent methods of calculation for inflationary predictions. For example, in the context of multi-field inflation, it is complicated to disentangle the gravitational and matter degrees of freedom when describing fluctuations produced in the  scalar fields using traditional methods. Typically, approximations are made to make the problem tractable which ignore back-reaction, that is, the effects of these very fluctuations on the background spacetime and fields trajectory. Restoring or even assessing the importance of these neglected effects then becomes extremely non-trivial. 
	
For instance, in \cite{Levasseur2010}, we presented an analytical study of mode coupling in the context of hybrid inflation. More precisely, we have studied the effects of short wavelength fluctuations 
both on the background inflaton and on long wavelength modes of the inflaton and waterfall fields. However, we found that our results were strongly dependent on initial solution chosen, as well as on the choice of background.
Because of this sensitivity, a more thorough study of the background was necessary, a goal which can be achieved via a study of stochastic inflation \cite{Starobinsky1984, Starobinsky1986, Goncharov:1987, Nakao:1988yi, Nambu:1988je, Nambu:1989uf, Habib1992, Linde:1993xx, Starobinsky1994}. Indeed, in \cite{Martin2011}, it was shown that stochastic effects can significantly alter the inflationary background dynamics in the context of hybrid inflation. 
 
 The stochastic inflation formalism is very powerful when it comes to addressing such issues, because it allows for a continual renormalisation of the background dynamics. 
	Its strength lies in its ability to separate the dynamics of coarse-grained long, classical, super-Hubble wavelengths from short, quantum fluctuation-dominated sub-Hubble wavelengths in a way that constantly corrects the background dynamics as modes of fluctuations are stretched from the quantum regime into the coarse-grained effective theory. The theory then describes the effective classical dynamics of the large-scale degrees of freedom, in the presence of a ``bath" where all the quantum fluctuations are collected in a classical noise term. Rather than being simply a computational trick, this feature is claimed to be fundamental to the system under study \cite{Starobinsky1986}. One of the major advantages of this formalism is then that, the super-Hubble theory being classical, one can solve it non-perturbatively, or to arbitrary order in perturbation theory \cite{Tsamis2005}, whereas it is unlikely one will ever be able to solve exactly the Heisenberg field equations for four dimensional interacting quantum fields.


	
The resulting theory shares a lot of similarities with quantum Brownian motion and out-of-equilibrium quantum statistical mechanics. Indeed, the classical system consisting of the super-Hubble modes is in contact with a bath of quantum fluctuations, modelled as a random noise source which must be averaged over time. The statistical properties of the noise are then all that is needed to derive the probability distribution of the system's state over many realizations.

This link with out-of-equilibrium mechanics suggests that one can derive the formalism to study inflationary dynamics from the machinery usually applied to quantum Brownian motion, such as the influence functional formalism \cite{Altland2010}. It arises in the context of the Schwinger-Keldysh closed time path (CPT) formalism, also known as the in-in formalism, which has become widely applied in cosmology since \cite{Weinberg:2005vy, Maldacena:2002vr}, after the pioneering work of \cite{Calzetta1987, Jordan1986} in the 1980's. However, in the case of the influence functional formalism, the strategy is to split the degrees of freedom of the full quantum fields into momentum space through a window function, and perform the path integral over the small scale fluctuations.

In the present paper, we will first derive the stochastic formalism  in Section II starting from the equations of motion of the fields, as was done in the early work of 
\cite{Vilenkin1983a, Vilenkin1983b, Starobinsky1986, Bardeen1986, Rey1987, Goncharov:1987, Sasaki:1987gy, Nakao:1988yi, Nambu:1988je, Nambu:1989uf, Salopek1990, Salopek1991, Habib1992}. However, being very careful in the derivation, we obtain terms which are not usually kept and we see precisely at what level the standard approach is making crucial, and often inconsistent, approximations. With this understanding, we present a new method to consistently treat the problem of back-reaction and mode-coupling in the context of stochastic inflation.

The key issue is that, during inflation, the propagator of the bath degrees of freedom evolve in a background which has to be fixed in order to compute the amplitude of the noise sourcing the bath, coarse-grained degrees of freedom. However, this effect shifts, or renormalizes, precisely this same background, making the propagator shift again. That is, as we will see, the bath degrees of freedom depend on the physical, renormalized background, not on a fixed free value as has been assumed in past studies of stochastic inflation. This makes the problem of integrating out the bath degrees of freedom much more difficult. Indeed, one must now make sure to evaluate at the same order both the coarse-grained fields and the quantum fields. We propose a new, recursive method to achieve this precise goal at the end of Section II.

In Section III, we explore a more rigorous approach inspired by condensed matter physics and prove that both derivations yield the same result for the Langevin equations at linear order. As we will see, one of the major advantages of the so-call influence functional formalism is that it makes it very natural to include higher loop corrections to mode-couplings and back-reaction effects. With this in mind, we develop a natural perturbative expansion and use it to calculate the full one-loop corrected Langevin equations in section IV, which is, to the best of our knowledge, a new calculation in the context of inflationary (with a time-dependent window function) spacetimes (see, however \cite{Morikawa1986, Boyanovsky1995}).

With this machinery developed, we then have all the necessary tools to apply our new recursive method to the specific example of two-field hybrid inflation potentials. We shall do such thing in a second paper \cite{secondpaper}.


\par 

\section{Stochastic Formalism: a Heuristic Derivation}
We would like to describe the evolution of scalar fields minimally coupled to gravity in de Sitter space. This task is very difficult to perform accurately using semi-classical methods, since one would expect effects such as self-coupling and interactions to become more and more important as time passes and the phase space of modes which have frozen and which are growing outside of the Hubble horizon becomes correspondingly  larger and larger. To tackle this problem we take advantage of the similarity of the current system with quantum Brownian motion. That is, we wish to use the techniques of the well-established Schwinger-Keldysh theory for out-of-equilibrium quantum field theory to derive the stochastic nature of the classical limit, in the form of a Langevin-like equation with a non-linear system-bath coupling. The techniques used in this case can be very technical and can even obscure the physical meaning of the approximations made. Therefore, we first present a heuristic derivation from the equations of motion, before proceeding to the more formal quantum field theoretic path integral approach.

\label{subsec:heuristicderivation}
\subsection{Definitions and formalism}

We consider scalar fields minimally coupled to gravity and evolving in a spatially flat background Friedmann-Lemaitre-Robertson-Walker space with the metric:
\bea
	ds^2&=&-dt^2 +a^2(t)d\vec{x}^2=a^2(\tau)\left(-d\tau^2+d\vec{x}^2 \right)\,;
\eea
where the second equality links the conformal time $\tau$ with the cosmic time $t$ by $d\tau=dt/a(t)$. 
Derivatives of $a(t)$ give the Hubble parameter, $H$, and the deceleration parameter $q(t)$:
\be
	H^2(t)\equiv\left(\frac{\dot{a}}{a}\right)^2=\frac{1}{3M_p^2}\rho\,,\quad q(t)\equiv -\frac{a\ddot{a}}{\dot{a}^2}=-1-\frac{\dot{H}}{H^2}\,.
\ee
Here, we have related $H$ to the matter energy density represented by $\rho$ through the first Friedmann equation.

For inflation to occur, we require positive expansion ($H(t)>0$) with $-1\leq q(t)<0$ (where the lower bound comes from the weak energy condition). When the deceleration $q$ saturates the lower bound, $H$ is constant and we obtain a locally de Sitter geometry, in which $a(t)$ is exactly exponential. In what follows, we shall not fix the background geometry to be exact de Sitter space, and consider quasi-de Sitter space (in the sense that we allow for $\dot{H}\neq0$). Additionally, for the sake of clarity of the notation, we will consider an unperturbed homogeneous and isotropic space-time in the current paper, and postpone the inclusion of metric perturbations to the second paper of this series, where we implement our method in a concrete example.

We are interested in studying effects due to the presence of multiple fields during inflation, and will therefore introduce a formalism treating two scalar fields $\Phi$ and $\Psi$. Generalizing it to a larger number of matter fields should be straightforward. The matter sector of the action we are considering is described by:
\be
	S_{\mathrm{m}}=  \int dt\int d^3{\bf x} \sqrt{-g}\mathcal{L}_{\mathrm{m}}(\Phi,\, \Psi)\, ,
\ee
with
\be	
 \mathcal{L}_{\mathrm{m}}(\Phi,\, \Psi)=\left[-\frac{1}{2}(\partial \Phi)^2 -\frac{1}{2}(\partial \Psi)^2-V(\Phi,\, \Psi) \right]\,.
\ee
It will be useful to split the potential into mass terms and interaction terms:
\be
\label{thegeneralpotential}
V(\Phi,\, \Psi) =\frac{1}{2}m^2_\Phi \Phi^2+\frac{1}{2}m_\Psi^2 \Psi^2+V_{\mathrm{pert}}(\Phi, \Psi)\,.
\ee


\subsection{Coarse-grained and quantum fields}

In the spirit of the early work on the subject \cite{Vilenkin1983a, Vilenkin1983b, Starobinsky1986, Bardeen1986, Rey1987,  Goncharov:1987, Sasaki:1987gy, Nakao:1988yi, Nambu:1988je, Nambu:1989uf, Salopek1990, Salopek1991, Habib1992}, we derive here the stochastic formalism using the fields equations of motion as a starting point. The underlying idea is to study them in Fourier space and split the modes into sub-Hubble and super-Hubble. We will be interested in studying the effective classical dynamics of the large-scale fields by treating the sub-Hubble modes collectively (through a coarse-graining process which will be defined shortly) as a classical noise source. 

Our strategy is the following: we wish to consider the full quantum fields $\Psi$ and $\Phi$, and split them into a large (in both amplitude and scale) ``homogeneous" piece and an inhomogeneous piece representing small, high wavenumber, quantum fluctuations:
\be
\label{phipsisplitting}
	\Phi=\varphi+\phi_>\,, ~ \Psi=\chi+\psi_>\, ,
\ee
where, by ``homogeneous", we mean the full fields $\Phi$, $\Psi$, coarse-grained over some large enough volume. Strictly speaking, it is an oversimplification to think about the coarse-grained fields as classical and it would be more accurate to denote $\varphi$ and $\chi$ by quantum mean fields (after all, from what has been said, both the coarse-grained and inhomogeneous fields are quantum by nature). However, as we will see later, it turns out they can actually be identified with the classical mean fields, while $\phi_>$ and $\psi_>$ can be identified with the quantum fluctuation fields. Also, see \cite{Calzetta1987, Morikawa1989, Habib1992, Albrecht1992, Calzetta1995, Hu1995, Polarski1995, Calzetta1997, Matacz1997, Kiefer1998, Kiefer2006, Kiefer2007, Weenink2011} and references therein for a discussion of decoherence and the precise condition required for quantum fields to undergo classicalization.

The question of defining the coarse-graining volume in a precise manner might also seem non-trivial, and we will quickly discuss it in a moment. For now, let us simply note that choosing an invariant scale seems like a sensible thing to do (in the same way that it is crucial to choose an invariant ultraviolet regularization scheme in order to regularize, for example, scalar fields UV divergences in the context of general relativity).

To give us a hint of what this invariant scale should be chosen to be, consider a given fixed comoving wavenumber $k=|{\bf k}|$ of a free, massless field sitting in de Sitter space. For this mode, significant particle production starts to occur when $k<Ha(t)$. Moreover, for free, massless scalars in exact locally de Sitter space, the range of physical wavelengths from zero to a constant $H^{-1}$ will have a constant dynamical impact since they lie in an invariant range \cite{Tsamis2005} (provided we choose an invariant regularization prescription). These facts all point toward using the de Sitter length as a cutoff scale to separate the coarse-grained fields $\varphi$ and $\chi$ from the quantum fields $\phi_>$, $\psi_>$. However, since we won't be considering exact de Sitter space in the following, and won't be restricting ourselves to massless fields, we expect that the above will only be approximately true in the case under study here. Regardless, this still suggests a natural choice of physical separation for the coarse-grained fields ($\varphi$, $\chi$) and ultraviolet fields ($\phi_>$, $\psi_>$), so that, in terms of the Fourier modes $k$ of the fields,
\bea
	\phi_{>}\,,~\psi_{>}& \mathrm{correspond~to}& \quad k>H(t)a(t)\, ,\\
	\varphi\,, ~\chi & \mathrm{correspond~to}& \quad H(t)a(t)>k>0 \, ,
\eea

\subsection{Effective classical coarse-grained dynamics}

The equations of motion satisfied by the full fields are:

\bea
\label{fullQuantumfieldsEOMs}
	-\Box \Phi+m_\Phi^2\Phi=-V_{\mathrm{pert},\Phi}(\Phi, \Psi)\, ,\\
	-\Box\Psi+m_\Psi^2\Psi=-{V}_{\mathrm{pert},\Psi}(\Phi, \Psi)\, ,
\eea
where $-\Box=\partial_{tt}+3H\partial_t-{\nabla^2}/{a^2}$ is the d'Alembertian and the commas denote a partial derivative(s) with respect to the field(s) following it.
Inserting (\ref{phipsisplitting}) into (\ref{fullQuantumfieldsEOMs}) and assuming that the quantum fields are a small perturbation to the coarse grained field, i.e. $\varphi\gg \phi_{>}$ and $\chi\gg\psi_{>}$, we write:

\bea
\label{splitPhieom}
	&-\Box\varphi+m_\Phi^2\varphi+{V}_{_\mathrm{pert},\Phi}(\varphi, \chi)&\nonumber\\
	&+\left[-\Box\phi_>+m_\Phi^2\phi_>+{V}^{_\mathrm{pert}}_{,\Phi\Phi}(\varphi, \chi)\phi_{>}+{V}^{_\mathrm{pert}}_{,\Phi\Psi}(\varphi, \chi)\psi_{>}\right]&\nonumber\\
	&=-{V}^{_\mathrm{pert}}_{,\Phi\Phi\Psi}(\varphi, \chi)\phi_{>}\psi_>&\nonumber \\
	&-\frac{1}{2}{V}^{_\mathrm{pert}}_{,\Phi\Phi\Phi}(\varphi, \chi)\phi_{>}^2 -\frac{1}{2}{V}^{_\mathrm{pert}}_{,\Phi\Psi\Psi}(\varphi, \chi)\psi_{>}^2+...\, ,&
\eea
where the ellipses stand for terms of third order or greater in the quantum fields. We obtain a similar equation for $\Psi$. The second line of the above equation, that is, the one in brackets, corresponds to the linearized equation satisfied by the small quantum perturbations $\phi_>$, in the presence of the mean field $\varphi$ background. Subtracting it from the full equation, one might think that one would find the equation for the coarse-grained field. 

However, things are slightly more complicated since the cutoff in Fourier space defining the comoving wavenumber range over which the quantum field extends is time-dependent. To see this, we start by expanding $\phi_>$ and $\psi_>$ into creation and annihilation operators on a time-dependent background:
\bea
	&\phi_>({\bf x}, t)=\int \frac{d^3 {\bf k}}{(2\pi)^3} W_H (k,t)\left[ \phi_{{\bf k}} \hat{a}_{{\bf k}}e^{-i{\bf k}\cdot{\bf x}}+\phi_{{\bf k}}^* \hat{a}^\dagger_{{\bf k}}e^{i{\bf k}\cdot{\bf x}} \right]&\nonumber \, ,\\
	&\psi_>({\bf x}, t)=\int \frac{d^3 {\bf k}}{(2\pi)^3} W_H (k, t)\left[ \psi_{{\bf k}} \hat{b}_{{\bf k}}e^{-i{\bf k}\cdot{\bf x}}+\psi_{{\bf k}}^* \hat{b}^\dagger_{{\bf k}}e^{i{\bf k}\cdot{\bf x}} \right]&\nonumber \, , \\
\label{quantumphipsiexpanded1}
\eea
where the operators $\hat{a}_{{\bf k}}, ~\hat{a}_{{\bf k}}^\dagger$ and $ \hat{b}_{{\bf k}},~ \hat{b}_{{\bf k}}^\dagger$ obey the standard commutation relations.

Here $W_H (k, t)$ is a window function that filters only the high-$k$ modes and projects out the long-wavelength modes with $k\lesssim a(t) H(t)$. The simplest choice of window function is the theta function $\theta\left[k/(\epsilon aH)-1\right]$, where $\epsilon<1$ is some constant small enough to ensure we include all the modes with a genuinely oscillatory quantum nature in the two quantum fluctuation fields. However, as was shown in \cite{Winitzki2000}, a sharp cutoff in k-space might be somewhat pathological, even if one could argue that such a sharp cutoff is only the mathematical limit of a smooth filtering procedure. Indeed, it has been shown \cite{Casini1999} that it does not satisfy the same fundamental properties as the asymptotic limit of any of these smooth limiting processes, and therefore has the physical effect of biasing the correlations of the coarse-grained fields on the largest scales to spuriously high values.

Most importantly, when the quantum fields are treated as free, massless fields in de Sitter space, it is the only window function that will cause them to appear as white noise to the coarse-grained fields, which means the latter will behave as a so-called Markovian process \cite{Altland2010}. This means that at every step of its evolution, the system made of the coarse-grained fields will experience noise which has no memory of previous history. On the contrary, considering a well-behaved smooth window function gives rise to colored noise, and consequently non-Markovian processes. Even though their treatment is more challenging, they are more physically motivated since they correspond to a finite volume cutoff between quantum and coarse-grained fields in configuration space, rather than a complicated infinite volume in the case of a sharp cutoff. Moreover, wide classes of smooth window functions give rise to the same asymptotic colored noise, which means that for choices of ``good" window functions, physical quantities are independent of the window function. In fact, as was shown in \cite{Winitzki2000}, the final correlations of the coarse-grained fields are independent of the window function if it is chosen to be 1) spherically symmetric and ${\bf x}$-dependent only through the combination $|{\bf x}|/R$ (with $R=\left[\epsilon a(t) H(t)\right]^{-1}$, the length scale over which we want to coarse-grain), in such a way that $W_H$ is constrained to have the form $\tilde{W}_H({\bf x}, t)\sim R^{-3}\tilde{w}(|{\bf x}|/R)$ when represented in configuration space and 2) such that the function $\tilde{w}(s)$ is decreasing at least as $\sim s^{-6}$ starting at $s\sim1$.  

In what follows, however, we will be interested in isolating the effects coming from colored noises induced by self- and cross-correlations of the quantum fields themselves, rather than colored noises induced by the choice of a smooth window function. Therefore, we leave $W_H$ unspecified throughout this paper. 

With the question of how to define the window function clarified, we can move on to describe how to remove the linearized equations satisfied by the quantum fields $\phi_>$, $\chi_>$ on sub-Hubble scales from the equations of motion for the full quantum fields $\Phi$, $\Psi$. We can now see why a naive subtraction cannot give the right result. Indeed, because of the time-dependent comoving cutoff of the window function in Fourier space, as modes are evolving in time, $k$-modes keep on exiting the Hubble radius to join the mean, coarse-grained fields and in doing so escape the range of $k$-modes where the quantum fields are defined. Therefore, upon subtracting the linearized equation obeyed by $\varphi$ from the full equation (\ref{splitPhieom}), we obtain:
\bea
	&-\Box\varphi+m_\Phi^2\varphi+{V}_{_{\mathrm{pert}},\Phi}(\varphi, \chi)= \delta S_{\phi_>}& \nonumber\\
	&-{V}^{_\mathrm{pert}}_{,\Phi\Phi\Psi}(\varphi, \chi)\phi_{>}\psi_>&\nonumber \\
	&-\frac{1}{2}{V}^{_\mathrm{pert}}_{,\Phi\Phi\Phi}(\varphi, \chi)\phi_{>}^2 -\frac{1}{2}{V}^{_\mathrm{pert}}_{,\Phi\Psi\Psi}(\varphi, \chi)\psi_{>}^2+...\, .&
\eea
Here, $\delta S_{\phi_>}$ represents the effect of all the $k$ modes within $\delta t$ which join the mean, coarse-grained field and give it an impulse (in the next section, we will see that this noise term arises from collectively treating all small scale modes inside a path integral; to give a rapidly-oscillating classical noise term). We can obtain an expression for $\delta S_{\phi_>}$ by substituting the expansion (\ref{quantumphipsiexpanded1}) in the second line of (\ref{splitPhieom}) and noting that, since the $\phi_k$'s are the linearized Fourier coefficients of $\Phi$ for $k\ll aH$, mode by mode they must satisfy the linearized Fourier transform of the total equation of motion (that is, $\ddot{\phi}_k+3H\dot{\phi}_k-(k^2/a^2-m_{\Phi}^2-V_{_{\uint},\Phi\Phi})\phi_k=-{V}_{_{\mathrm{pert}},\Phi\Psi}\, \psi_k$). We find:
\bea
	\delta S_{\phi_>}=\int\frac{d^3{\bf k}}{(2\pi)^3}\left[ 3H\dot{W}_H\left(\frac{k}{\epsilon a(t)H(t)}\right)\hat{\phi}_{\bf k}\right.\nonumber\qquad \quad\\
	\left.+2\dot{W}_H\left(\frac{k}{\epsilon a(t)H(t)}\right)\dot{\hat{\phi}}_{\bf k}+\ddot{W}_H\left(\frac{k}{\epsilon a(t)H(t)}\right)\hat{\phi}_{\bf k}\right]e^{-i{\bf k}\cdot{\bf x}} \nonumber\\
\eea
where 
\bea
	\hat{\phi}_{\bf k}e^{-i{\bf k}\cdot{\bf x}}= \left[\phi_{{\bf k}} \hat{a}_{{\bf k}}+\phi_{-{\bf k}}^* \hat{a}^\dagger_{-{\bf k}} \right] e^{-i{\bf k}\cdot{\bf x}}\, .
\eea

We can see how this expression relates to the qualitative interpretation of $\delta S_{\phi_>}$ given above as the impulse of every Fourier mode coming from the quantum field $\phi_>$ when it joins the coarse-grained field by looking at the simplest example of window function, $W_H=\theta\left[k/(\epsilon a H)-1\right]$. Indeed, in this case, the time-derivative transforms the $W_H$ it into a $\delta$-function in the first two terms. This means that out of the integral over those two terms, only the mode crossing the window function at a given $t$ contributes, and therefore these terms represent the kick to the mean field coming from the mode with $k=\epsilon a(t)H(t)$. The last term includes the derivative of a $\delta$-function, which might seem unsettling. However (as would be seen if we had been working with the Hamiltonian first-order formalism rather than second-order equations of motion \cite{Salopek1991}), all these terms need to be included in order to represent the impulse to the coarse-grained field coming from both $\phi_k$ displacement {\it and} momentum.

From the coarse-grained field point of view, we see that the quantum fluctuations escaping the Hubble radius act as a sustained noise source. Indeed, the statistical properties of the $\phi_k$ modes becoming larger than the coarse-graining region and constantly sourcing $\varphi$ are that of a Gaussian-distributed (as the $\phi_k$'s satisfy the linearized Fourier space equation), zero-mean stochastic noise term $\xi^\phi({\bf x} , t)$, with a variance which can be calculated from the variance of the quantum field. Note that, at this point, one should technically interpret this stochastic random noise as the piece of the quantum fluctuations effectively acting as a classical field. As we will see in the next section, the piece remaining purely quantum will induce an extra term introducing a non-local mass renormalization of the coarse-grained field, as well as dissipation. However, for the purpose of interest here where the slow-roll conditions are satisfied, this term is negligible and we will not include it in the following.

Defining
\bea
	&\xi^\phi_1= -\int\frac{d^3{\bf k}}{(2\pi)^3} \dot{W}_H\left(\frac{k}{\epsilon a(t)H(t)}\right) \left[\phi_{{\bf k}} \hat{a}_{{\bf k}}e^{-i{\bf k}\cdot{\bf x}}+\phi_{{\bf k}}^* \hat{a}^\dagger_{{\bf k}}e^{i{\bf k}\cdot{\bf x}} \right],& \nonumber\\
	&\xi^\phi_2=\int\frac{d^3{\bf k}}{(2\pi)^3} \dot{W}_H\left(\frac{k}{\epsilon a(t)H(t)}\right) \left[\dot{\phi}_{{\bf k}} \hat{a}_{{\bf k}}e^{-i{\bf k}\cdot{\bf x}}+\dot{\phi}_{{\bf k}}^* \hat{a}^\dagger_{{\bf k}}e^{i{\bf k}\cdot{\bf x}} \right],& \nonumber \\
\label{noiseformforphi}
\eea
it is straightforward to show that the noise term induced by the linear $\phi_>$ quantum modes exiting the Hubble radius is given by:
\be
	\delta S_{\phi_>}=3H\xi^\phi_1+\dot{\xi}^\phi_1-\xi^\phi_2 \, .
\ee
Defining $\xi^\psi({\bf x}, t)$ in a similar fashion for the $\Psi$ field, we can obtain an identical expression for $\delta S_{\psi_>}$. 

Note that here, it is strictly the time-dependence of the window function rather than mode-coupling (as would be the case in standard quantum Brownian motion) which gives rise to noise. Indeed, the noise only represents the continuous Hubble-crossing of the modes, rather than some fundamental coupling between super- and sub-Hubble modes. This means that the properties of the noise terms generated in this way are such that they will not give rise to decoherence (in technical terms, a mass term in the equations of motions can only change the squeezing of a state, not induce and increase of its phase-space area or entropy, which means it cannot lead to its classicalization). See \cite{Habib1992, Weenink2011} for an in-depth discussion of these points.

In the case of the potential (\ref{thegeneralpotential}), we obtain the following equations of motion up to second order in the quantum fields:
\bea
	&\Box{\varphi}+m_\Phi^2\varphi+{V}_{_{\mathrm{pert}},\Phi}\,\varphi =3H\xi^\phi_1+\dot{\xi}^\phi_1-\xi^\phi_2&\nonumber\\
	&-{V}_{_{\mathrm{pert}},\Phi\Phi\Psi}(\varphi, \chi)\phi_{>}\psi_>&\nonumber \\
	&-\frac{1}{2}{V}_{_{\mathrm{pert}},\Phi\Phi\Phi}(\varphi, \chi)\phi_{>}^2 -\frac{1}{2}{V}_{_{\mathrm{pert}},\Phi\Psi\Psi}(\varphi, \chi)\psi_{>}^2\, ,&
\label{stochastichybridfulleomphi}
\eea
\bea
	&\Box \chi+m_\psi^2\chi+{V}_{_{\mathrm{pert}},\Psi}\,\chi=3H\xi^\psi_1+\dot{\xi}^\psi_1-\xi^\psi_2&\nonumber\\
	&-{V}_{_{\mathrm{pert}},\Psi\Psi\Phi}(\varphi, \chi)\psi_{>}\phi_>&\nonumber \\
	&-\frac{1}{2}V_{_{\mathrm{pert}},\Psi\Psi\Psi}(\varphi, \chi)\psi_{>}^2 -\frac{1}{2}{V}_{_{\mathrm{pert}},\Psi\Phi\Phi}(\varphi, \chi)\phi_{>}^2 \, .&
\label{stochastichybridfulleompsi}
\eea
The non-linear terms on the right-hand-side of both equations represent couplings between the coarse-grained fields and the small-scale quantum fluctuations. Therefore, one way of picturing the situation is to think about the coarse-grained fields as a system of long-ranged fluctuations immersed in a bath of small scale quantum fluctuations. The latter will back-react on the system and cause its decoherence, triggering the transition from long-range quantum mean fields to mean, coarse-grained classical fields. In other words, the quantum mean fields become dressed by quantum fluctuations and they decohere through interactions with these quantum fluctuations around them, developing random {\it classical} fluctuations in the process \cite{Calzetta1995}. Note, however, that classicalization is not {\it only} caused by inflation self-interactions. In fact, {all} non-linear couplings will contribute to this process (see, e.g., a treatment of classicalization from non-linear couplings to the gravitational tensor or vector modes \cite{Calzetta1995, Franco2010}, or \cite{Weenink2011} for an exploration of other possibilities).

Equations (\ref{stochastichybridfulleomphi}) and (\ref{stochastichybridfulleompsi}) form a system of {\it classical} Langevin equations, sourced by random gaussian noise terms (which are completely determined by their two-point correlation functions). They describe a stochastic process, which means that rather than solving for one realization of $\varphi$ and $\chi$, we must solve for their probability distribution $\rho(t,\varphi, \chi)$ over many realizations, through a Fokker-Planck equation \cite{Starobinsky1994}. Expectation values of functionals of the stochastic fields, in particular their correlation functions, can then be calculated via:
\be
	\langle F[\varphi(t, {\bf x}), \chi(t, {\bf x})]\rangle=\int\int d\omega_\varphi d \omega_\chi\rho(t, \omega_\varphi, \omega_\chi)F(\omega_\varphi, \omega_\chi)\, ,
\ee
where $\omega_{\phi,\psi}$ are simply dummy variables for the fields.

For completeness, the linearized equations of motion satisfied by the Fourier components of the quantum fields $\phi_k$ and $\psi_k$ are:
\bea
	\ddot{\phi}_k+3H\dot{\phi}_k+\left( \frac{k^2}{a^2}+m_\Phi^2+V_{_{\mathrm{pert}},\Phi\Phi}  \right)\phi_k&&\nonumber\\
\label{phiquantumlinearizednoisehybrid}
	=-{V}_{_{\mathrm{pert}},\Phi\Psi}\, \psi_k\qquad \qquad&&\\
	\ddot{\psi}_k+3H\dot{\psi}_k+\left(\frac{k^2}{a^2}+m_{\Psi}^2+{V}_{_{\mathrm{pert}},\Psi\Psi} \right)\psi_k&&\nonumber \\
	=-{V}_{_{\mathrm{pert}},\Psi\Phi}\,\phi_k\qquad \qquad&&
\label{psiquantumlinearizednoisehybrid}
\eea

From this, we can calculate the statistical properties of $\xi^{\phi, \psi}_{1, 2}$. Solving the full system (\ref{phiquantumlinearizednoisehybrid}), (\ref{psiquantumlinearizednoisehybrid}) in the presence of the classical coarse-grained fields (\ref{stochastichybridfulleomphi}), (\ref{stochastichybridfulleompsi}) is highly non-trivial, since not only is it a non-linear system with time-dependent effective mass for the two quantum fields, but the mean fields themselves are shifted by the presence of quantum fluctuations. In order to make the problem tractable, it is conventional to fix the quantum fields to be free, massless fields in exact de Sitter space, and to take their statistical properties to be simply defined by the Bunch-Davis vacuum \cite{Rey1987, Salopek1991, Tsamis2005, Martin2011}. This approximation is justified at first order; however we are interested in studying effects such as the tilt of the spectrum and the strength of back-reaction effects of small scale modes on large scale fluctuations; precisely the effects that require a more precise knowledge of the details of the modes behavior coming from the quantum fields when they join the coarse-grained fields in order to be self-consistent. 

This is in contrast with most previous works who have been claiming to find non-perturbative solutions for correlations of the inflaton (e.g. \cite{Geshnizjani2005, Geshnizjani2005a, Kunze:2006tu, Finelli:2008zg}), by which they mean that they are solving the classical coarse-grained system (\ref{stochastichybridfulleomphi}), (\ref{stochastichybridfulleompsi}) to arbitrary order in perturbation theory (in terms of the slow-roll parameters), in the presence of a noise $\xi$ which is calculated from fixing the quantum fields to their exact Bunch-Davis values in de Sitter space. This approach of including higher orders in $\varepsilon$ in the classical coarse-grained theory, whilst only using leading order approximations (in the slow-roll parameters) for the noise sourcing the classical perturbations is a priori inconsistent. In most cases of single field inflation, in particular for chaotic inflation, this nuance is just a technical point which doesn't really have measurable consequences, since the main source of non-gaussianities and the tilt in the spectrum are effects incorporated in the classical theory. Indeed, the tilt in $\langle\delta \varphi,\, \delta \varphi \rangle$ will then come from the factor of $H$ in (\ref{stochastichybridfulleomphi}) and (\ref{stochastichybridfulleompsi}), and the non-gaussianities from the derivative of the potential evaluated over the classical fields which is allowed to receive corrections proportional to the slow-roll parameters through the time-variation of the background $\varphi$ as it rolls down the potential. As shown in, e.g., \cite{Finelli:2008zg, Finelli:2010sh}, one then recovers agreement between the ``naive" stochastic method and the regular perturbative approach. In particular, it has been shown \cite{Tsamis2005} that for a scalar field with an arbitrary potential, one can in this way correctly capture all leading order logarithms $[\ln (a)]^n$ of coincident correlators to all orders in perturbation theory.

However, if two fields are present, situations might arise where it would be crucial that a correction to $\xi$ is included. In particular, it would be the case in the scenario where one field is heavy or even develops a tachyonic instability, as, for example, is the case in the final stages of hybrid inflation, where the value of the quantum fields correlators as they cross the window function deviate significantly from their Bunch-Davis values. Even more importantly, the approach outlined above neglects the presence of back-reaction from the quantum fields onto the coarse-grained fields by setting the interactions between the two sectors of the theory to zero. In what follows, we will outline a tentative process to remedy to these flaws.

\subsection{Outline of the expansion strategy}

Our goal is to define a process which gives an increasingly precise result by including consistently every contribution at each order in the slow-roll parameters and in the fields. At zeroth order we start by solving the purely classical, homogeneous, noiseless equations of motion under the slow-roll approximation. To zeroth order in slow-roll parameters, the small quantum fields are then approximated by their free, massless, de Sitter estimates, as was done in the previous studies mentioned above. This makes finding the variance of the classical noise terms $\xi^{\phi, \psi}_{1, 2}$ from the two point correlation function of $\phi_k$ and $\psi_k$ possible, since in this case (\ref{phiquantumlinearizednoisehybrid}) and (\ref{psiquantumlinearizednoisehybrid}) can be solved exactly in terms of the usual Hankel functions. Combined with the sharp cutoff choice for the window function, i.e. the theta function $\theta\left[k/(\epsilon aH)-1\right]$, we then find that the stochastic terms $\xi^{\phi, \psi}_{1, 2}$ are exactly random, Gaussian, {\it white} noise terms, as expected.

We can then substitute these values of the classical stochastic noise in (\ref{stochastichybridfulleomphi}) and (\ref{stochastichybridfulleompsi}) (keeping only terms of order lower or equal in slow-roll to that of the square root of the noise variance, the noise having zero mean by definition). Moreover, for the sake of consistency, since the stochastic noise incorporates exclusively the linearized quantum mode functions, we expand the coarse-grained equations to first order in the field perturbations in all other terms. This means all back-reaction terms from the small scale quantum fields onto the coarse-grained fields will be neglected at this order. The final resulting coarse-grained evolution equation we find is then a Markovian process.  We can solve this corrected equation for the statistical properties of the coarse-grained fields. In particular, we are interested in finding expressions for the mean value of $\varphi$ and $\chi$ over many realizations, as well as their variance and more generally the whole probability density functions $\rho(t,\varphi, \chi)$. Effectively, we can think about this new classical background as being shifted from its zeroth order classical (or ``bare") value by the small-scale quantum fluctuations constantly exiting the Hubble radius, freezing out, and joining the coarse-grained field.

Qualitatively, it is easy to understand that, from the point of view of a quantum fluctuation deep inside the Hubble radius during inflation (more precisely, a $k$-mode well above the cutoff scale), the background it feels it is evolving in is not the ``bare", zeroth order background. Rather it is the collective background composed of the bare value plus all the quantum modes which already have frozen out and have become classical. This shifted background is exactly the one which the method outlined above made precise and allowed us to solve for in the previous step of our method. Because of its stochastic nature, this solution is under the form of a probability distribution of the coarse-grained fields. This motivates the next step in our method, which is then to go back to (\ref{phiquantumlinearizednoisehybrid}) and (\ref{psiquantumlinearizednoisehybrid}), and replace all occurrences of the coarse-grained fields by their average values, variances and higher momenta (this has been proposed, e.g. in \cite{Abolhasani2011, Guth2012} and more recently in \cite{Lazzari2013}):
\bea
	\varphi, \chi\qquad&\rightarrow&\qquad \langle\varphi\rangle, \langle\chi\rangle,\nonumber\\
	\varphi^2, \chi^2 \qquad&\rightarrow&\qquad \langle\varphi^2\rangle, \langle\chi^2\rangle,\\
	{\varphi^p\chi^q \qquad}&{\rightarrow}& {\qquad \langle\varphi^p\chi^q\rangle\, ,}
\eea
and solve the corrected linearized equations for the small quantum fluctuation fields, this time expanding to next-to-leading order in slow-roll (since we want to pick up the effects of the statistically shifted coarse-grained fields on the mass of the perturbations; in other words, we now solve for the full linearized mode functions). Qualitatively, this corresponds to re-expanding to find the behaviour of quantum fluctuations around the adjusted background. From the results obtained for these corrected quantum fields, we can find the corresponding statistical properties of corrected stochastic noise term, which will now be colored rather than white. More precisely, it will not be uniformly random, since it's maximal amplitude will be time-dependent, but it will again give rise to a Markovian process in the sense that the noise will still be memoryless.
 
From there, we can go back to the coarse-grained system (\ref{stochastichybridfulleomphi})-(\ref{stochastichybridfulleompsi}), and reevaluate the probability distribution and the correlation functions of the classical fields in the presence of the corrected linearized noise, which now includes non-zero correlations. To do this in a consistent manner, we now keep in the coarse-grained equations terms up to order $\sqrt{\langle \xi^{\phi,\psi}_{1,2}\,\xi^{\phi,\psi}_{1,2}\rangle}$ in slow-roll. This will allow us to find the spectrum of fluctuations (by solving for the variance of the classical coarse-grained associated field) predicted by stochastic inflation, including the tilt, in a consistent manner. 

This self-consistent implementation is expected to have important consequences, in cases where one or both of the fields are not test fields, in the sense that the noise amplitude depends on the stochastic value of the coarse-grained field itself - that is, when the amplitude of the noise is not (or cannot be approximated as being) time-independent$^1$\let\thefootnote\relax\footnote{$^1$In this case, one needs to assign a {so called ``prescription''} $\alpha \in [0,1]$ to the Langevin equation in question, {which sets at which point $t+\alpha\mathrm{d}t$ the noise must be calulated when the field is incremented between $t$ and $t+ \mathrm{d}t$, when defining the Langevin dynamics as a limit of a discrete stochastic pocess,} and the Fokker-Planck equation associated with this Langevin equation will depend explicitly on $\alpha$. We thank Vincent Vennin for pointing this out, and we plan to return to this point in a later study.}.

Two specific cases where we can expect stochastic effects to play an interesting role when compared with results previously obtained using the stochastic formalism are 1) in the neighborhood of an instability point in the potential, and 2) the case where the secondary field has a mass that is parametrically larger than the inflaton, such as in the case of quasi-single field inflation, or in hybrid inflation. In particular, in the case of hybrid inflation, we expect stochastic effects to play a preponderant role in driving the evolution of the background during the waterfall phase \cite{Martin2011}. We derive the precise form of these effects in the second paper of this series \cite{secondpaper}.

Moreover, at this order in slow roll, we can now study the effect of back-reaction coming from second-order effects in the quantum fields. In particular, we can estimate the amplitude of the second-order terms in the quantum fields in (\ref{stochastichybridfulleomphi})-(\ref{stochastichybridfulleompsi}) and see if back-reaction can have a significant effect on the coarse-grained fields. In the context of hybrid inflation, we are particularly interested in studying this effect during the waterfall, when the $\psi_>$ field becomes tachyonic and the $\psi_k$ modes enter a regime of tachyonic resonance. In this case, we expect the collective effect of the exponentially growing modes to leave an imprint in the background. If we find that the second-order terms become dominant over the linearized stochastic noise, this would be indicate a breakdown of our perturbative approach in the quantum fields, which in turn can be used to rule out models. Alternatively, we can estimate the second-order contribution of the quantum field to the stochastic noise in the presence of the stochastically-shifted background, which can then be used as a criteria to measure the break down of perturbation theory. We will come back to this study in a later paper.

\section{Schwinger-Keldysh derivation}
\label{subsec:SKderiv}
In order to define more precisely the idea of mean classical fields arising from purely quantum fields, as well as to understand to what extend the primordial fluctuations are quantum and at which point they can be understood as providing a stochastic noise source to the background, we need to turn to the more technical formalism of out-of-equilibrium quantum field theory.

\subsection{CPT formalism}

We once again start by splitting the full quantum fields $\Phi$ and $\Psi$ into a long-wavelength and a short-wavelength pieces, with the cutoff between the two again defined at a scale around the Hubble radius. As before, we also define the small-wavelength part to be:
\bea
	&\phi_>({\bf x}, t)=\int \frac{d^3 {\bf k}}{(2\pi)^3} W_H (k,t)\left[ \phi_{{\bf k}} \hat{a}_{{\bf k}}e^{-i{\bf k}\cdot{\bf x}}+\phi_{{\bf k}}^* \hat{a}^\dagger_{{\bf k}}e^{i{\bf k}\cdot{\bf x}} \right]&\nonumber\\
	&\psi_>({\bf x}, t)=\int \frac{d^3 {\bf k}}{(2\pi)^3} W_H (k, t)\left[ \psi_{{\bf k}} \hat{b}_{{\bf k}}e^{-i{\bf k}\cdot{\bf x}}+\psi_{{\bf k}}^* \hat{b}^\dagger_{{\bf k}}e^{i{\bf k}\cdot{\bf x}} \right]&\nonumber \\
\label{quantumphipsiexpanded}
\eea
where $W_H$ is still the same window function projecting out the long-wavelength modes which we will, again, pick to be a theta function in what follows. The mode functions $\phi_{\bf k}$, $\psi_{\bf k}$ are chosen to be solutions of the linearized field equations, a point which we will make more specific later.

The long-wavelength part of the fields will then simply be given by $\varphi=\Phi-\phi_>$ and $\chi=\Psi-\psi_>$, again as before. Substituting this decomposition in the action for the scalar fields, we see that it splits into two distinct actions for each part of the fields, plus an interaction piece:

\bea
	S[\varphi, \chi, \phi_>, \psi_>]&=&S_{\usys}[\varphi, \chi]+S_{\ubath}[ \phi_>, \psi_>]+\nonumber \\
	&&\qquad S_{\uint}[\varphi, \chi, \phi_>, \psi_>] \,.
\eea

The fields $\varphi$ and $\chi$ represent the so-called {\it system} which we want to study, in the presence of $\phi_>$ and $\psi_>$ which constitute the {\it bath}. In the following, our goal will be to perform an integration of the action over the bath degrees of freedom (the short wavelength fields), $\phi_>$ and $\psi_>$, in order to obtain an effective action $S_{\ueff}$ for the system of long-wavelength modes, and finally take the classical limit. 


In order to obtain the in-in expectation values for out-of-equilibrium quantum field theory as opposed to the usual in-out transition amplitude, we proceed using the Closed Time Path (CPT) formalism \cite{Schwinger1961, Feynman1963}. In this framework, time integrals are made along closed contours starting at an initial time $t_i$ far in the past, evolving forward in time up to some time $t_0$ and then backward to the initial, distant time. The initial states are prepared in this distant past, point at which interactions and effects of curvature are assumed to vanish. This way, the states can be set to be their free vacuum state in Minkowski space, that is, the Bunch-Davis vacuum. We then split each field in the theory into two sets of fields, the first set for fields living on the forward piece of the contour, denoted by a ``$^+$", and the second one by ``$^-$" for fields living on the backward piece. This means $\Phi$ is now decomposed into 4 scalar fields: $\varphi^+$, $\varphi^-$, $\phi_>^+$, and $\phi_>^-$, and similarly for $\Psi$.

\begin{widetext}

Since we want to integrate over the fields $\phi_>^+$, $\phi_>^-$, $\psi_>^+$, and $\psi_>^-$, the object we are interested in is the reduced density matrix $\hat{\rho}_r$ and its time-evolution operator $\mathcal{J}_r$. From the total density matrix $\hat{\rho}$, we define the reduced density matrix $\hat{\rho}_r$ as:
\be
	\hat{\rho}_r(\varphi^+, \chi^+, \varphi^-, \chi^- )=\int_{-\infty}^{\infty} d\phi_>^+  d\psi_>^+ \int_{-\infty}^{\infty}  d\phi_>^- d\psi_>^-\hat{\rho}(\varphi^+, \chi^+, \phi_>^+, \psi_>^+\,;\, \varphi^-, \chi^-, \phi_>^-, \psi_>^-)\delta(\phi_>^+-\phi_>^-)\delta(\psi_>^+-\psi_>^-)
\ee
and is propagated in time from $t_i$ to $t_0$ by the reduced evolution operator $\mathcal{J}_r$:
\be
	\hat{\rho}_r(\varphi^+_f, \chi^+_f, \varphi^-_f, \chi^- _f, t_0) =\int_{-\infty}^{\infty} d\varphi^+_i d\chi^+_i \int_{-\infty}^{\infty}d\varphi^-_i d\chi^-_i \mathcal{J}_r(\varphi^+_f, \chi^+_f, \varphi^-_f, \chi^-_f, t_0~|~\varphi^+_i, \chi^+_i, \varphi_i^-, \chi_i^-, t_i) \hat{\rho}_r(\varphi^+_i, \chi^+_i, \varphi^-_i, \chi^-_i, t_i )\, .
\ee
Here, the index $f$ denotes the boundary values of the fields at the time $t_0$. We would like to write $\mathcal{J}_r$ in a path integral representation. This is in principle possible, however, in practice the explicit expression can be very complicated since $\mathcal{J}_r$ depends on the initial state. The crucial simplifying assumption we make is that at very early times $t_i$, the system and the bath are not correlated, that is:
\be
\hat{\rho}(t=t_i)=\hat{\rho}_{\usys} (t_i)\times \hat{\rho}_{\ubath}(t_i)\, ;
\ee
in which case the evolution operator $\mathcal{J}_r$ does not depend on the initial state of the system. This is a reasonable assumption in our case, since at sufficiently early times, the modes of the bath were deep inside the Hubble horizon and felt as if free in flat, Minkowski space. 

We can then write the reduced evolution operator in terms of a functional representation, and doing so we find an expression for the effective action for the system:

\bea
	\mathcal{J}_r(\varphi^\pm_f, \chi^\pm_f, t_0~|~\varphi^\pm_i, \chi^\pm_i, t_i)	&=&
	\int^{\varphi^\pm_f}_{\varphi^\pm_i} \mathcal{D}\varphi^\pm\int^{\chi^\pm_f}_{\chi^\pm_i}\mathcal{D} \chi^\pm \exp\left( \frac{i}{\hbar}\left\lbrace S_{\usys}[\varphi^+, \chi^+]-S_{\usys}[\varphi^-, \chi^-]\right\rbrace\right)F[\varphi^\pm, \chi^\pm] \\
	& \equiv & \int^{\varphi^\pm_f}_{\varphi^\pm_i} \mathcal{D}\varphi^\pm\int^{\chi^\pm_f}_{\chi^\pm_i}\mathcal{D} \chi^\pm \exp\left\lbrace\frac{i}{\hbar} S_{\ueff}[\varphi^\pm, \chi^\pm]\right\rbrace\, , \qquad \qquad \qquad\qquad \qquad \qquad\,\,
\eea
where $F[\varphi^\pm, \chi^\pm]$ is known as the {\it influence functional}, produced from the action of the bath on the system. In general, it is a non-local, highly non-trivial object. Not only does it depend on the time history, but it also mixes the forward and backward histories along the time path in an irreducible manner. In our case of particular initial conditions, it is given by:
\bea
	F[\varphi^\pm, \chi^\pm] =  \int_{-\infty}^{\infty}d\phi_{>_f}^+d\psi_{>_f}^+ \int_{-\infty}^{\infty}d\phi_{>_i}^\pm d\psi_{>_i}^\pm 
	\int_{\phi_{>_i}^\pm}^{\phi_{>_f}^+} \mathcal{D} \phi_{>}^\pm  \int_{\psi_{>_i}^\pm}^{\psi_{>_f}^+} \mathcal{D} \psi_{>}^\pm \qquad \qquad\qquad\qquad\qquad \qquad \qquad\qquad  \qquad\qquad\nonumber \\ 
	\exp\left(\frac{i}{\hbar} \left\lbrace S^0_{\ubath}[ \phi^+_>, \psi^+_>]-S^0_{\ubath}[ \phi_>^-, \psi_>^-]+S_{\uint}[\varphi^+, \chi^+, \phi_>^+, \psi_>^+] -S_{\mathrm{pert}}[\varphi^-, \chi^-, \phi_>^-, \psi_>^-]\right\rbrace\right)\hat{\rho}_{\ubath}(\phi_{>_i}^\pm,\psi_{>_i}^\pm)\, , \nonumber\\
	 {\equiv}  \exp\left[\frac{i}{\hbar}S_{\uIA}[\varphi^\pm, \chi^\pm] \right]\, .\quad\qquad\qquad \qquad \qquad\qquad \qquad\qquad \qquad \qquad\qquad ~~\, \qquad\qquad \qquad  \qquad \qquad\qquad
	\label{definitionoftheinfluenceaction}
 \eea
Here we have used $S^0_{\ubath}$ to denote the action of the free bath degrees of freedom. Because of our choice of initial Bunch-Davis conditions, the integral of the initial density matrix over initial conditions of the bath fields can be assimilated to an overall constant (which we won't write in the following to avoid cluttering the notation). Note also that we have imposed the boundary conditions $\phi_{>_f}^+=\phi_{>_f}^-$ and $\psi_{>_f}^+=\psi_{>_f}^-$ at $t=t_0$. In the last line, we have defined $S_{\uIA}$, which is called the {\it influence action}. Its relation to the effective action is easily found to be: $S_{\ueff}[\varphi^\pm, \chi^\pm]=S_{\usys}[\varphi^+, \chi^+]-S_{\usys}[\varphi^-, \chi^-]+S_{\uIA}[\varphi^\pm, \chi^\pm]\equiv S_{\usys}^+-S_{\usys}^-+S_{\uIA}$. We can write the influence functional as:
\be
\label{inflienceactionlong}
\exp\left\lbrace\frac{i}{\hbar}S_{\uIA}[\tilde \varphi, \tilde \chi] \right\rbrace=\qquad \qquad \qquad \qquad \qquad \qquad \qquad \qquad \qquad\qquad \qquad \qquad \qquad \qquad \qquad \qquad \qquad  \qquad \qquad \qquad 
\ee
\bea
	\int_{-\infty}^{\infty}d\phi_{>_f}^+ d\psi_{>_f}^+ 
	\int^{{\phi_{>_f}^+}} \mathcal{D} \phi_{>}^{\pm} \int^{{\psi_{>_f}^+}}\mathcal{D}\psi_{>}^{\pm} e^{\frac{i}{\hbar}\int d^4x\left[\left( \frac{1}{2} \tilde{\phi} ^T_>\tilde{\Lambda}_\phi \tilde{\phi}_>  + \tilde{\varphi} ^T\tilde{\Lambda}_\phi \tilde{\phi}_>  \right)+\left( \frac{1}{2}\tilde{\psi} ^T_>\tilde{\Lambda}_\psi\tilde{ \psi}_>+ \tilde{\chi} ^T\tilde{ \Lambda}_\psi \tilde{ \psi}_>  \right)- \tilde{V}_{\ubath}(\tilde \phi_{>}, \tilde \psi_>) -\tilde{V}_{\uint}(\tilde\varphi, \tilde\chi, \tilde\phi_>, \tilde\psi_>) \right]}\, ,\nonumber
\eea
where the time integral in the above spans from $-\infty$ to $t_0$, the turn-around time in the CTP contour$^1$\let\thefootnote\relax\footnote{$^1$However, since we impose the boundary conditions $\Phi^+({\bf x},t_0)=\Phi^-({\bf x},t_0)$ and $\Psi^+({\bf x},t_0)=\Psi^-({\bf x},t_0)$, we could easily extend the upper bound of integration to future infinity, as the forward and backward parts of the contour cancel exactly beyond $t_0$. Indeed, $t_0$ being the time at with we are evaluating a particular operator, causality of the theory requires this cancellation.}, and where we have introduced the matrix notation for simplicity:
\bea
&\tilde{\phi}_>=\left( \begin{array}{c} \phi_>^+ \\ \phi_>^- \end{array} \right) \qquad \tilde{\psi}_>=\left( \begin{array}{c} \psi_>^+ \\ \psi_>^- \end{array} \right)\qquad \tilde{\varphi}=\left( \begin{array}{c} \varphi^+ \\ \varphi^- \end{array} \right) \qquad \tilde{\chi}=\left( \begin{array}{c} \chi^+ \\ \chi_>^- \end{array} \right)& \nonumber \\
&\tilde{\Lambda}_\phi =\left( \begin{array}{cc} \Lambda_\phi & 0 \\ 0 & -\Lambda_\phi \end{array} \right) \qquad \tilde{\Lambda}_\psi =\left( \begin{array}{cc} \Lambda_\psi & 0 \\ 0 & -\Lambda_\psi \end{array} \right)\, , &
\label{matrixnotationshort}
\eea
with the superscript $T$ meaning transpose, and we have also introduced $\Lambda_\phi$ and $\Lambda_\chi$, which are defined to be the integration kernels for the free scalar fields:
\be
	\Lambda_\phi= -a^3(t)\left[\partial^2_t+3H\partial_t-\frac{\nabla^2}{a^2(t)} +m^2_\Phi \right]\, ; \qquad \qquad \Lambda_\psi= -a^3(t)\left[\partial^2_t+3H\partial_t-\frac{\nabla^2}{a^2(t)} +m_\Psi^2 \right]\, .
	\label{integrationkernelsforscalarfields}
\ee

In the first two terms in the exponential we put all the terms which are bilinear in the fields, i.e. the terms over which we can compute the path integral exactly to obtain the leading order influence action, and in the last two terms we put all non-linear couplings, for which in general we need to do a perturbative expansion in the coupling constants in order to perform the path integral over the bath degrees of freedom. Note that $\tilde{V}$ with any subscript are the vector form of terms which were part of $V_{\mathrm{pert}}$ as was defined in the previous section, as all potential terms in $\tilde{V}$ involve more than two-fields interactions. In the case of the generic potential (\ref{thegeneralpotential}), this will generate both loop corrections coming from the self-interactions of the baths degrees of freedom, represented above by the term $\tilde{V}_{\ubath}$, and loop corrections due to the interactions of the bath with the system, represented by the term $\tilde{V}_{\uint}$.

\subsection{Schwinger-Keldysh derivation of the Langevin equations}

We want to start by calculating the leading order influence action $S_{\uIA}^{(1)}$ in order to obtain the propagators (or equations of motion) for the system degrees of freedom once the leading interactions between bath and system have been integrated out (note that at zeroth order, the influence functional trivially only adds a multiplicative factor to the system path integral). Therefore, at this order, we neglect the interacting potential both in the bath and at the bath-system level beyond the terms linear in the interactions, and only perform the exact path integral over bilinear terms in (\ref{inflienceactionlong}), as done in detail in Appendix A. That is, only taking into account linear and quadratic terms in the influence action, we obtain a gaussian integral which can be performed to yield:
\bea
\label{firstorderinfluencefunctional1}
	S_{\uIA}^{(1)}= \frac{i}{2\hbar}\int d^4x d^4x'\varphi_q(x)\mathrm{Re}\left[ \Pi_\phi(x, x') \right]\varphi_q(x')-\frac{2}{\hbar}\int d^4 xd^4x'\theta(t-t')\varphi_q \mathrm{Im} \left[ \Pi_\phi(x, x') \right]\varphi_c+\left(\chi\leftrightarrow \varphi \right)\, ,
\eea 
where:
\be
\label{Pioperator}
	 \Pi_\phi(x, x') = \int \frac{d^3{\bf k}}{(2\pi)^3} a^3(t)\left[P_t\phi_{{\bf k}}(t)\right] e^{-i{\bf k}\cdot{\bf x}}a^3(t')\left[ P_{t'}\phi^*_{{\bf k}}(t')\right]e^{i{\bf k}\cdot{\bf x'}}\,,
\ee
with $P_t$ defined as:
\be
\label{ptoperator}
P_t=\left[ \ddot{W}_H(t)+3H\dot{W}_H(t)+2\dot{W}_H(t)\partial_t \right]\, ;
\ee
and where we have moved to the so-called Keldysh basis and have defined the classical and quantum fields: $\varphi_c$, $\chi_c$ and $\varphi_q$, $\chi_q$, respectively, by \cite{Altland2010}:
\be
	\left(\begin{array}{c}\varphi_c \\ \varphi_q \end{array} \right)\equiv\left( \begin{array}{c} \frac{\varphi^++\varphi^-}{2} \\ \varphi^+-\varphi^- \end{array}\right)\, , \qquad 
	\left(\begin{array}{c}\chi_c \\ \chi_q \end{array} \right)\equiv\left( \begin{array}{c} \frac{\chi^++\chi^-}{2} \\ \chi^+-\chi^- \end{array}\right).
\label{transfotoclassquant}
\ee
\end{widetext}

In this manner, we reproduce the result of \cite{Morikawa1990}. Note however that here the mode functions appearing in $\Pi_{\phi, \psi}(x,x')$ are technically calculated from solutions to the full massive mode equations of motion, since we considered the masses to be part of the tree-level propagators as opposed to perturbation. This will be of importance later, when we include interaction and obtain system fields as part of the mode function equation for the bath degrees of freedom, since in that case the mode function appearing in similar expressions will again be calculated from the {\it full} mode function equation of motion, with the actual system fields (as a justification for this, recall that all the operations done here are inside a path integral for the system fields), rather than on some fixed artificial background.

The leading order of the influence action which we defined in (\ref{definitionoftheinfluenceaction}) by $ F[\varphi^\pm, \chi^\pm]=\exp\left[\frac{i}{\hbar}S_{\uIA}^{(1)}\right]$ therefore splits into a real and imaginary part. As pointed out for example in \cite{Caldeira1983, Grabert1988}, they respectively represent noise and dissipation in the system fields. Indeed, the kernels Im$\left[\Pi_\phi(x, x')\right]$ and Im$\left[\Pi_\psi(x, x')\right]$ are known as the dissipation kernels for each of the classical fields. The motivation for the name comes from the fact that their non-symmetric part each introduces an extra term in the equation of motion of each classical field when taking the variational derivative of the effective action, $\delta S_{\ueff}/ \delta \varphi_q$ and $\delta S_{\ueff}/ \delta \chi_q$. These terms are non-local and proportional to $\dot{\varphi}_c$ and $\dot{\chi}_c$, respectively, and can therefore be assimilated into friction, or in other words dissipation. However, as long as we are interested in studying the equations of motion while the slow-roll approximation is valid, these friction terms will be negligible in comparison to the Hubble friction terms, related to the Hubble expansion (as shown, for example, in \cite{Morikawa1990, Matarrese2004}).

Similarly, the symmetric part of the kernels Im$\left[\Pi_\phi(x, x')\right]$ and Im$\left[\Pi_\psi(x, x')\right]$ each give rise to a mass-renormalisation term which does not contribute to the mixing of the forward and backward histories \cite{Hu1992}. These terms can also be shown to be negligible \cite{Morikawa1990, Matarrese2004}. Therefore, in the following, we will neglect the contributions to the effective action $S_{\ueff}$ proportional to Im$\left[\Pi_\phi(x, x')\right]$ and Im$\left[\Pi_\psi(x, x')\right]$ altogether$^1$.

On the other hand, the terms containing the real part of $\Pi_\phi$ and $\Pi_\psi$ in (\ref{firstorderinfluencefunctional1}) give an imaginary part to the effective action. As such, this term cannot be properly interpreted as an ordinary part of the action. That is, as detailed in Appendix A, this terms can be reinterpreted as a result of a weighted average over the configurations of stochastical noise terms representing the sub-Hubble quantum fluctuations which couple to $\varphi$ and $\chi$. To make this explicit, it is possible to re-introduce these as classical noise terms with the appropriately chosen probability distribution \cite{Morikawa1986}, as is customary to do in out-of-equilibrium condensed matter physics or when studying Brownian motion \cite{Altland2010}. We will do just this in the following. 

Before we do so, we first comment on how this interpretation of the real and imaginary part of the influence action can be linked with known results from out-of-equilibrium statistical mechanics. As in the theory of quantum Brownian motion, the presence of quantum fluctuations gives rise to stochastic effects in the system through two distinct effects: the first one is a rapidly-oscillating and stochastically distributed noise term in the systems fields equations of motion; the second one proportional to the velocity of the system fields is a friction term describing how fast the system approaches equilibrium from an out-of-equilibrium configuration. The latter depends on the correlation of the stochastic force, and is thus not independent from the noise term. In out-of-equilibrium statistical mechanics, this relation is known as the {\it fluctuation-dissipation theorem}. It is evident here because the dissipation and noise terms are related through the real and imaginary parts of the same kernel. 
Typically, in non-equilibrium statical mechanics, this theorem further states that the ratio of the transport coefficient and the microscopic random fluctuations is proportional to the temperature of the system. One could therefore ask whether in our case the ratio of the two terms we find here can be related to the Hawking temperature $H/2\pi$ of de Sitter space. As shown in \cite{Morikawa1990}, it is indeed possible to define a `temperature' of order of the Hawking temperature, but (in the case of quasi-de Sitter space) which depends on $\nu$ and $\epsilon$, and therefore is a property specific to the matter-spacetime interaction.

%

\begin{widetext}
\let\thefootnote\relax\footnote{$^1$In particular, \cite{Morikawa1990} argued these terms to be $\mathcal{O}(\epsilon^3)$, whereas the Re$\left[\Pi_{\phi,\psi}(x, x')\right]$ terms we keep here are at most $\mathcal{O}(\epsilon^2)$.}
Here, in order to be able to properly interpret the imaginary part of the influence functional as noise, we introduce two real classical random fields per field in the system $\xi_1^{\phi}$, $\xi_2^\phi$ and  $\xi_1^{\psi}$, $\xi_2^\psi$, each obeying the Gaussian probability distribution:
\be
\label{noiseprobdistribution}
	\mathcal{P}\left[\xi_1^{\phi,\psi}, \xi^{\phi,\psi}_2 \right]= \exp\left\lbrace -\frac{1}{2}\int d^4xd^4x' [\xi^{\phi,\psi}_1 (x), \xi^{\phi,\psi}_2(x)] {\bf A}^{-1}(x,x')\left[\begin{array}{c}\xi^{\phi,\psi}_1(x')\\  \xi^{\phi,\psi}_2(x') \end{array}\right] \right\rbrace\, ,
\ee
where we have defined, letting $r=|{\bf x}-{\bf x'}|$ {and $x=\left(\tau,{\bf x}\right)$},
\be
\label{themonstercorrelationmatrix}
{\bf A}^{i,j}_\phi(x, x')=\frac{H^2}{2\pi^2}\int \frac{d k}{k}  \frac{\sin (kr)}{kr} k^5 \tau \tau' \partial_{k\tau}W_H\left(k\tau\right)\partial_{k\tau'}W_H \left(k\tau'\right)\left[ 1+\frac{\dot H({\tau})}{H^2({\tau})} \right]\left[ 1+\frac{\dot H({\tau}')}{H^2({\tau}')} \right]\mathrm{Re}\left[ {\bf M}_\phi^{i,j}(k\tau, k\tau')\right]\, ,
\ee
where
\be
\label{theMmatrices}
	 {\bf M}_\phi^{i,j}(k\tau, k\tau')=\left(\begin{array}{cc} \phi_{{\bf k}}(t) \phi^*_{{\bf k}}(t')  & q\left(-k\tau\right) \phi_{{\bf k}}(t) \phi^*_{{\bf k}}(t') \\  \phi_{{\bf k}}(t) \phi^*_{{\bf k}}(t')q\left(-k\tau'\right)& q\left(-k\tau\right) \phi_{{\bf k}}(t) \phi^*_{{\bf k}}(t') q\left(-k\tau'\right)  \end{array}\right)\, ,
\ee
and similarly for the matrices ${\bf A}^{i,j}_\psi(x, x')$ and $ {\bf M}_\psi^{i,j}(k\tau, k\tau')$. The functions $q_{\phi,\psi}(-k\tau)$ represent the $k$-dependent (and time-dependent only through the combination $k/aH$) part of the logarithm derivative of the mode function of the quantum fields, as defined in (\ref{thePfunctiontointbyparts}). Also, note that, again, the mode functions appearing in ${\bf M}_\psi^{i,j}(k\tau, k\tau')$ are calculated from the full mode functions equations of motion.

After some manipulations (see Appendix A), we can now use the influence functional to rewrite the leading-order effective action. To do so, we recall that we are interested in taking the classical limit of the effective action. As mentioned above, the fields $\varphi_q$ and $\chi_q$ embody the quantum nature of the difference between the configurations on the forward and backward pieces of the time contour. In order to explore the classical limit of the effective action, we proceed to the field rescaling $\varphi_q,\,\chi_q\rightarrow \hbar\varphi_q,\, \hbar\chi_q $  \cite{Altland2010}. With this redefinition, a first order expansion of the effective action in $\hbar\varphi_q,\, \hbar\chi_q $ corresponds to the $\hbar$-independent, or classical, sector of the action:

\bea 
 	\int^{\varphi^+_f}_{\varphi^\pm_i} \mathcal{D}\varphi^\pm \int^{\chi^+_f}_{\chi^\pm_i}\mathcal{D} \chi^\pm  \exp\left[\frac{i}{\hbar} S^{(1)}_{\ueff}\right] \qquad\qquad\qquad \qquad\qquad\qquad \qquad\qquad\qquad \qquad\qquad\qquad \qquad\qquad\qquad\qquad \nonumber\\
	=\int \mathcal{D}\varphi^{q,c}\mathcal{D} \chi^{q,c} \int\mathcal{D}\xi_1\mathcal{D}\xi_2 \mathcal{P}\left[\xi_1, \xi_2 \right] \exp\left( i\int d^4x a^3(t)\left\lbrace\varphi_q\left[\left(\Box-m^2_{\Phi}\right) \varphi_c -{V}_{_{\mathrm{pert}},\Phi}\left(\varphi_c, \chi_c\right) \right] \nonumber \right. \right. \qquad\qquad\\
	\left.\left. +\chi_q\left[\left(\Box-m_\Psi^2\right)\chi_c -{V}_{_{\mathrm{pert}},\Psi}\left(\varphi_c, \chi_c\right) \right] +\varphi_q\left[ p_\phi(t)\xi^\phi_1+\xi^\phi_2\right]-\dot{\varphi}_q\xi^\phi_1+\chi_q\left[ p_\psi(t)\xi^\psi_1+\xi^\psi_2\right]-\dot{\chi}_q\xi^\psi_1
	  \right\rbrace\right) \,,
\label{classicalfirstorderaction}
\eea
\end{widetext}
where $p_{\phi,\psi}$ are defined as the $k$-independent part of the derivative of the mode function of the quantum fields. Note that the separation of the influence action (\ref{firstorderinfluencefunctional1}) we have just done between a weight part $\mathcal{P}\left[\xi_1, \xi_2 \right]$ and a pure action part contributing to the effective action $S^{(1)}_{\ueff}$ is unique \cite{Morikawa1986}. The fields $\xi_1^{\phi,\psi}$ and $\xi_2^{\phi,\psi}$ can then easily be interpreted as classical random fields, since they obey Gaussian statistics and their weight function has a real exponent.

The equations of motion for the classical fields $\varphi_c$ and $\chi_c$ in the classical limit are then given by:
\be
	\left. \frac{\delta S^{(1)}_{\ueff}}{\delta \varphi_q}\right|_{\varphi_q=0}=0~~;\quad\left. \frac{\delta S^{(1)}_{\ueff}}{\delta \chi_q}\right|_{\chi_q=0}=0.
\label{langevinequationssaddlepointcondition}
\ee
Note that we evaluate these for $\varphi_q=\chi_q=0$ since our choice of boundary conditions at $t=t_0$ enforce that the actual time axis is unique. 
Relabeling $t_0\rightarrow t$ for simplicity, we therefore obtain:
\bea
\label{stochasticphifromCTP}
	(-\Box+m_\Phi^2)\varphi_c +\tilde{V}_{,\Phi}(\varphi_c, \chi_c)&=&\nonumber\\
	 p_\phi(t)\xi^\phi_1&+\xi^\phi_2&+\dot{\xi}_1^\phi+3H\xi^\phi_1,\\
 	(-\Box +m_\Psi^2)\chi_c+ \tilde{V}_{,\Psi}(\varphi_c, \chi_c)&=&\nonumber\\
	p_\psi(t)\xi^\psi_1&+\xi^\psi_2&+\dot{\xi}_1^\psi+3H\xi^\psi_1.
\label{stochasticpsifromCTP}
\eea
Here, the Gaussian noises are defined to have zero mean, and therefore their statistical properties are completely determined by the matrices ${\bf A}^{i,j}_\phi(x, x')$ and ${\bf A}^{i,j}_\psi(x, x')$. In particular there can (and will) be cross-correlations between $\xi_1^{\phi,\psi}$ and $\xi_2^{\phi,\psi}$, and their self- and cross-correlation functions will depend both on the choice of window function and the level of precision we use to specify the mode functions (and their derivative through the $q_{\phi,\psi} \left(-k\tau\right)$ functions introduced in (\ref{theMmatrices})) at Hubble-crossing. This is of crucial importance for the implementation of the solution method we outlined at the end of the section \ref{subsec:heuristicderivation}.

Comparing the result we just obtained in (\ref{stochasticphifromCTP}) and (\ref{stochasticpsifromCTP}) with (\ref{stochastichybridfulleomphi}) and (\ref{stochastichybridfulleompsi}), we find that the stochastic terms are exactly consistent, since if we go back to the way we defined $\xi_{1~\uold}^{\phi,\psi}$ and $\xi_{2~\uold}^{\phi,\psi}$ in section \ref{subsec:heuristicderivation}, we have:
\bea
	\xi^\phi_{1~\uold}&=&\int \frac{d^3 {\bf k}}{(2\pi)^3}\dot{W}_H \hat{\phi}_{\bf k}e^{-i{\bf k}\cdot{\bf x}}\, ,\\
	\xi^\phi_{2~\uold}&=& \int \frac{d^3 {\bf k}}{(2\pi)^3}\dot{W}_H \dot{\hat{\phi}}_{\bf k}e^{-i{\bf k}\cdot{\bf x}}\, ,
\eea
and similarly for $\xi^\psi_2$. However, recall that, along the way of the derivation of the stochastic equations of motion we presented in Appendix A, we assumed the time derivative of the mode functions of the quantum fields, $\phi_k$ and $\psi_k$, could be written as:
\be
\label{eq:defpq}
	 \dot{\hat{\phi}}_{\bf k}=\left[p_\phi(t) +q_{\phi} \left(-k\tau\right)\right]\hat{\phi}_{\bf k}.
\ee
We therefore find that $\xi^\phi_{1~\uold}$ and $\xi^\psi_{2~\uold}$ can be split into:
\bea
   \xi^\phi_{1~\uold}&=&\xi^\phi_{1~\unew}=\xi^\phi_{1}\, ,\\
	\xi^\phi_{2~\uold} &=&-p_\phi(t) \xi^{\phi}_1-\xi_{2~\unew}^\phi,
\eea
and similarly for $\xi^\psi_{1~\uold}$ and $\xi^\psi_{2~\uold}$, where, to obtain the second term of $\xi^{\phi,\psi}_{2~\uold}$ in this form, we have absorbed the $q_{\phi, \psi} \left(-k\tau\right)$ functions in the matrices ${\bf A}^{i,j}_\phi(x, x')$ and ${\bf A}^{i,j}_\psi(x, x')$, as we have done above, and where we have used that $\dot{W}_H \approx H k\tau \left( 1+\dot{H}/H^2 \right)\partial_{k\tau}W_H$.

We therefore obtain the perfect consistency between the two methods, up to linear order in the bath fields. In addition, it is very easy to extend the above method using the usual tools from field theory to include higher order (in the bath degrees of freedom) corrections to (\ref{stochasticphifromCTP}) and (\ref{stochasticpsifromCTP}), and understand effects such as back-reaction.

We now want to understand how this calculation can be done in the presence of a time-dependent window function (see e.g. \cite{Morikawa1986,Hu1993, Boyanovsky1995} for a calculation of loop corrections up to second order in the context of a single scalar field with a power-law potential in Minkowski space and a time-independent window function), and see how these corrections give rise to the back-reaction terms we found in (\ref{stochastichybridfulleomphi}) and (\ref{stochastichybridfulleompsi}).

\bigskip
\section{One-Loop Corrections}
\label{subsec:OneLoop} 
\subsection{Setup}

In order to illustrate how loop corrections can be understood and calculated, it will be useful to assume a particular form of interaction potential between bath and system degrees of freedom, $\tilde{V}_{\uint}(\tilde{\varphi}, \tilde{\chi}, \tilde{\phi}_>, \tilde{\psi}_>)$. In particular, this boils down to choosing an interacting potential for the fields $\Phi$ and $\Psi$, which  we choose a renormalizable, $ \mathbb{Z}_2$-symmetric interacting potential in the symmetry phase, as follows:

\bea
\label{sampleIntPot}
	{V}_{\mathrm{pert}}(\Phi, \Psi)= \frac{\lambda}{4!}\Psi^4+\frac{g^2}{2}\Phi^2\Psi^2 \, .
\eea

It will also be interesting to consider the case of a mass hierarchy between the two fields: $m_\Phi\ll m_\Psi$. 

The first thing to do here is then to split the full fields into bath and system degrees of freedom, as we have done above, and to expand perturbatively in the bath fields, with the goal of performing their path integral order by order in a loop expansion, in a fashion very similar to what is done in Wilsonian renormalization \cite{Wilson:1973jj,Wilson:1974mb}.

But, before we do so, at this point we need to define the free action for the bath degrees of freedom. Indeed, we are facing something which looks like an ambiguity here, because once (\ref{sampleIntPot}) is expanded in system and bath fields, $\varphi$, $\chi$ and $\phi_>$, $\psi_>$, respectively, it gives rise to interaction terms which are bilinear in the bath degrees of freedom, but contains further powers of the system fields, such as e.g. a $\frac{\lambda}{4}\chi^2 \psi^2_>$ term. We need to decide whether these terms shall be included in the propagators of the free bath fields as time-dependent masses, or treated as perturbations, as explained in \cite{Boyanovsky1995}. Happily, it is possible to show that these two methods are equivalent, as will be explained later and in Appendix C. For now, let us simply argue that, physically, quantum fluctuations which are part of the bath (and so sub-Hubble), are evolving on a background which they feel to be the zero mode plus all modes which have already crossed the Hubble radius and which we are now collectively treating as the coarse grained fields, even at tree level. It is therefore intuitive to include terms which are bilinear in the bath fields such as $\frac{\lambda}{4}\chi^2 \psi^2_>$ as time-dependent masses in the free propagator of the bath degrees of freedom.

Finally, let us also stress that one should really use the {\it actual} system field in the bath propagator, since we are doing the path integral over the bath fields {\it inside} the path integral over the system fields. Another way to see that this is indeed the case would be to recast this problem in the language of the 2PI formalism \cite{Millington2012, Berges2004}, which we leave to a further study.

\begin{widetext}
With these issues clarified, and going back for a moment to the $+$/$-$ notation for the fields rather than the Keldysh basis for the classical/quantum notation, we can now move on to writing down the free bath and interacting action corresponding to (\ref{sampleIntPot}) for the $+$ and $-$ parts of the contour:
\bea
	S^{0+}_{\ubath}-S^{0-}_{\ubath}+S_{\uint}^\pm&=&\int dx^4  \left.{\Bigg(}\frac{1}{2} \tilde{\phi}_>^T\tilde\Lambda^{(2)}_\phi\tilde{\phi}_> +\tilde{\varphi}^T\tilde\Lambda^{(1)}_\phi\tilde{\phi}_>+ \frac{1}{2} \tilde{\psi}_>^T\tilde\Lambda^{(2)}_\psi\tilde{\psi}_> +\tilde{\chi}^T\tilde\Lambda^{(1)}_\psi\tilde{\psi}_> -\right.\nonumber\\
	&a^3(t)&\left\lbrace\left[2g^2\varphi^+\chi^+\phi^+_>\psi^+_> +g^2\left(\varphi^+\right)^2\phi_>^+\left(\psi^+_>\right)^2 +g^2\left(\chi^+\right)^2\psi_>^+\left(\phi^+_>\right)^2+\frac{\lambda}{3!}\chi^+(\psi^+_>)^3\right]-\left\{+\rightarrow- \right\} \right\rbrace \nonumber \\
	&+&\left.\mathcal{O}\left[(\psi_>^\pm)^4, (\phi_>^\pm\psi_>^\pm)^2\right]\right.\Bigg)
\label{perturbativeintaction}
\eea
where we have introduced a matrix notation similar to the one introduced in (\ref{matrixnotationshort}), with:
\bea
&\tilde{\Lambda}^{(1),(2)}_{\phi, \psi} =\left( \begin{array}{cc} \Lambda^{+(1),(2)}_{\phi,\psi} & 0 \\ 0 & -\Lambda^{-(1),(2)}_{\phi,\psi}  \end{array} \right)\, ,  &
\label{matrixnotationshort2}
\eea
and the integration kernels to first and second order in the bath fields are defined to be: 
\bea
&\Lambda^{\pm,(1)}_\phi=\Lambda^{\pm,(2)}_\phi=a^3(t)\left[ \Box-m_\Phi^2-g^2\left(\chi^\pm\right)^2\right]\,,&\nonumber\\
&\Lambda^{\pm,(1)}_\psi=a^3(t)\left[ \Box -m_\Psi^2-g^2\left(\varphi^\pm\right)^2-\frac{\lambda}{3!}\left(\chi^\pm\right)^3\right]\, , \qquad \Lambda^{\pm,(2)}_\psi=a^3(t)\left[\Box-m_\Psi^2-g^2\left(\varphi^\pm\right)^2-\frac{\lambda}{2}\left(\chi^\pm \right)^2\right]&\, .
\label{lambda2}
\eea
\end{widetext}

\subsection{Unperturbed influence functional and free propagators}
In order to obtain the loop expansion and calculate the effective potential for the system degrees of freedom once the bath fields have been integrated out, 
we introduce four currents, one for each field on the forward part of the time contour, and one per field on the backward part, as in the standard procedure \cite{Calzetta1987, Hu1993}. We then proceed as outlined in Appendix B to define a perturbative expansion of the corresponding influence functional. 

Evaluating the first term in this perturbative series yields the unperturbed influence functional $F^{(1)}[\tilde{J}_\phi, \tilde{J}_\psi]$, as defined in (\ref{unperturbedinfluencefunctional}), which can be explicitly calculated:
\bea
\label{unperturbedinfluencefunctionalTEXT}
	F^{(1)}[\tilde{J}_\phi, \tilde{J}_\psi]=\exp\left[  \ln \Det \left(\tilde{\Lambda}^{(2)}_\phi\right)^{-1/2}\right. \qquad\qquad\\
	\left. -\frac{i}{2\hbar}\int d^4x d^4x'\tilde{J}^T_\phi \left(\tilde{\Lambda}^{(2)}_\phi\right)^{-1}\tilde{J}_\phi + (\phi \rightarrow \psi)\right] \, .\nonumber
\eea
Note that here, $ \left(\tilde{\Lambda}^{(2)}_\phi\right)^{-1}$ can be defined in terms of Green's functions on the contour, in analogy to (\ref{firstdefofprops}). Also, because of our definition of the free propagators for the bath degrees of freedom, the determinant terms in the above are not simply an extra multiplicative constant added to the influence functional, and cannot be neglected as is sometimes assumed.

From here, it is easy to verify that the one point functions of the bath fields gives zero, that is, 
\bea
\langle \phi^\pm_> (x)\rangle_0=\langle \psi^\pm_> (x)\rangle_0=0\, .
\eea
Furthermore, we can compute the two-point functions:
\bea
\label{pertpropforphi}
	\langle\tilde{\phi}_>(x)\tilde{\phi}_>^T(y)\rangle_0\qquad\qquad\qquad \qquad \qquad \qquad \qquad \qquad \\
	=\left( \begin{array}{c c} T\langle\hat \phi_>^+(x)\hat \phi_>^+(y)\rangle  & \langle\hat \phi_>^-(y)\hat \phi_>^+(x)\rangle\\ \langle\hat \phi_>^-(x)\hat \phi_>^+(y)\rangle & \bar{T}\langle\hat \phi_>^-(x)\hat\phi_>^-(y)\rangle \end{array}\right)\, ,\nonumber\\
	= i\hbar\left(\tilde{\Lambda}^{(2)}_\phi\right)^{-1}\nonumber \, ,\quad\qquad\qquad \qquad \qquad \qquad \,\,
\eea
and similarly for $\tilde{\psi}_>$. As expected, we also find $\langle\tilde{\phi}_>(x)\tilde{\psi}_>(y)\rangle=0$. 

Here to calculate these correlators we need to expand the fields in Fourier space in terms of creation and annihilation operators. The mode functions appearing then need to be solutions to the equations of motion $\Lambda^{(2)}_{\phi}(k)\phi_{{\bf k}}=0$, $\Lambda^{(2)}_{\psi}(k)\psi_{{\bf k}}=0$, where $\Lambda^{(2)}_{\phi, \psi}(k)$ are defined as the Fourier transform of the kernels defined in (\ref{matrixnotationshort2})-(\ref{lambda2}). 

Now, since we are interested in writing down a diagrammatic expansion, the above correlators are interpreted as the bath propagators $i\hbar\Delta_{\phi}(x,y)$. More specifically, for fields time-ordered along the forward CPT contour, they are equivalent to the standard time-ordered Feynman propagator, $i\hbar\Delta_F(x,y)$, while for anti-time-ordered fields along the backward part of the contour they correspond to the anti-time ordered Dyson propagator $i\hbar\Delta_D(x,y)$. The upper off-diagonal element of (\ref{pertpropforphi}) represents bath fields propagating from the $+$ to the $-$ pieces of the contour, and therefore has $x^0$ always `earlier' than $y^0$. As such, it is proportional to the absolutely-ordered negative-frequency Wightman propagator $i\hbar\Delta_<(x,y)$. Similarly, the lower off-diagonal element of (\ref{pertpropforphi}) represents bath fields propagating from the $-$ to the $+$ pieces of the contour, and analogously is proportional to the positive-frequency Wightman propagator $i\hbar\Delta_>(x,y)$. Therefore, we can identify:
\bea
	\langle\tilde{\phi}_>(x)\tilde{\phi}_>^T(y)\rangle\equiv i\hbar\Delta_{\phi}(x,y)=\qquad \qquad \qquad \qquad \qquad \\
	i\hbar \left( \begin{array}{c c} \Delta_{\phi,F}(x,y)  & -\Delta_{\phi,<}(x,y) \\ -\Delta_{\phi,>}(x,y) & \Delta_{\phi,D}(x,y) \end{array}\right)\, ,\nonumber
\eea
These propagators are represented below:

\begin{figure}[h]
\begin{center}
\includegraphics[width=8cm]{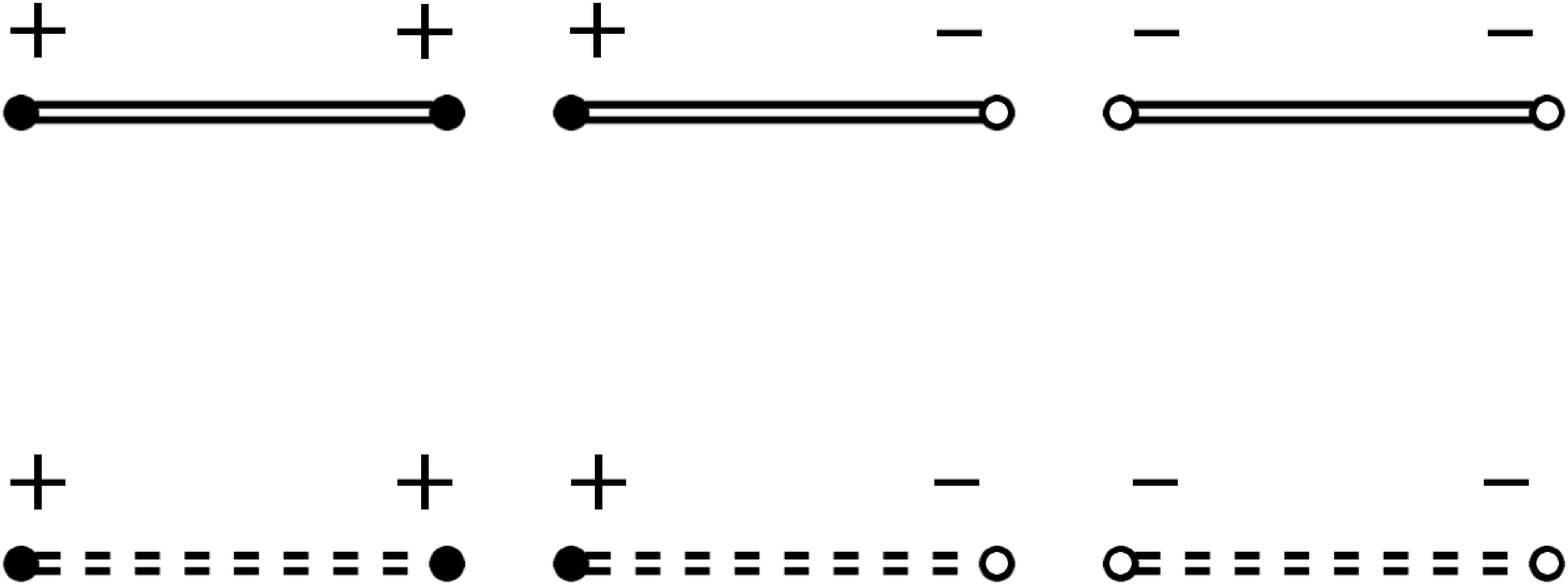}
\label{fig:freeProps}
\end{center}
\end{figure}
%
%
%
We use double lines to represent bath degree of freedom, and reserve the use of single lines for system fields. Dashed lines (either double or single) are used to represent $\Phi$ fields, while plain lines are used to represent $\Psi$ fields. Moreover, in the following we will use solid dots to represent vertices on the forward piece of the contour, while open dots will be used to represent vertices on the backward piece of the contour. Similarly, the signs `$+$' and `$-$' indicate if the field represented sits on the forward or backward piece of the contour. To simplify the notation, for bath fields we will only indicate `$+$' or `$-$' on the appropriate line, while for system external legs we write explicitly which field the line belong to.

From here, converting the bath fields propagators to the Keldysh basis, where the fields are expressed in terms of classical and quantum fields, we find:
\bea
	i\hbar\tilde \Delta_{\phi}(x,y) =\qquad\qquad\qquad\qquad\qquad\qquad\qquad\qquad\qquad\nonumber\\
	\nonumber \left( \begin{array}{cc} 4\mathrm{Re}\left[ \Omega_\phi(x, y)\right] & 2i\Theta(y^0-x^0)\mathrm{Im} \left[\Omega_\phi(y, x)\right] \\ 2i\Theta (x^0-y^0)\mathrm{Im} \left[\Omega_\phi(x, y)\right]&  0  \end{array} \right)
\eea
\be
\ee
where
\bea
\label{eq:Omega}
\Omega_\phi(x,y) =\int \frac{d^3k}{(2\pi)^3}W^t_{H} W^{t'}_H\phi_{\bf k}(x^0)\phi^*_{\bf k}(y^0)e^{-i{\bf k}\cdot({\bf x}-{\bf y})} \, ,
\eea
and similarly for $\tilde{\psi}_>$, with $\Omega_\psi(x,y)$ analogously defined.

\subsection{One-loop calculation}

Having found an expression for the bath fields propagators, we can move on to computing the remaining one-loop contributions. Note that diagrams involving vertices with only bath degrees of freedom will give their first contribution at two loops, and so we conclude that terms involving $\tilde{V}_{\ubath}(\phi_>^\pm, \psi_>^\pm)$ will give no contribution at one loop. The same goes for terms $\mathcal{O}(\psi_>^4, \psi_>^2\phi_>^2)$. From (\ref{perturbativeintaction}) we see that the first contribution to the loop expansion (see (\ref{perturbativeloopexpansion}) for the explicit details of the expansion) will come from terms involving two powers of the interacting action:
\bea
\label{eq:F2order}
	F[\varphi^\pm, \chi^\pm]\supset
	\frac{i}{\hbar} \left[ \frac{i}{2\hbar}\left\langle \tilde{V}^2_{\uint}\left(\varphi^+, \chi^+, \phi_>^+, \psi_>^+\right) \right\rangle_0\right.\nonumber\\
	 +\frac{i}{2\hbar}\left\langle \tilde{V}^2_{\uint}\left(\varphi^-, \chi^-, \phi_>^-, \psi_>^-\right) \right\rangle_0\qquad \qquad\nonumber\\
	 \left.-\frac{i}{\hbar}\left\langle \tilde{V}_{\uint}(\varphi^+, \chi^+, \phi_>^+, \psi_>^+)\tilde{V}_{\uint}(\varphi^-, \chi^-, \phi_>^-, \psi_>^-) \right\rangle_0\right]\nonumber \, ,
\eea
coming from the following terms in the action:
\bea
\label{premierTermeQuad}
	&\int dx^4  \left[\tilde{\varphi}^T\tilde\Lambda^{(1)}_\phi\tilde{\phi}_>+\tilde{\chi}^T\tilde\Lambda^{(1)}_\psi\tilde{\psi}_> \right]\, ,&\\
	&-\int dx^4 a^3(t)\left[2g^2\varphi^+\chi^+\phi^+_>\psi^+_> +\left\{ +\rightarrow -\right\}\right]\, .&
\label{deuxiemetermQuad}
\eea
We now work out the impact of these two terms one after the other. As before, one would argue that the two terms in (\ref{premierTermeQuad}), which are linear terms in the bath degrees of freedom, vanish, since the elements of the pairs $\{ \varphi, \phi\}$ and $\{\chi, \psi\}$ are defined to be orthogonal to each other. However, in our case, the kinetic term in $\tilde\Lambda^{(1)}_{\phi, \psi}$ acts on the window function in $\tilde{\phi}_>$ and $\tilde{\psi}_>$, making the contribution from these terms non-vanishing. This is the exact effect we calculated throughout section \ref{subsec:SKderiv}, and with no surprise we find their contribution to be
\bea
	-\frac{i}{2\hbar}\int d^4xd^4x' \left[a^3(t)\tilde\varphi^T_x\vec{\square}_x \sigma_3\Delta_\phi(x,x')\sigma_3\cev{\square}_{x'}\tilde{\varphi}_{x'}a^3(t')\right.\nonumber\\
	\left.+\{ \varphi \rightarrow \chi,~\phi \rightarrow \psi\}\right]\, ,\qquad \qquad \qquad
\eea
where $\sigma_3$  is the third Pauli matrix. Transforming to the Keldysh basis, and proceeding through a series of manipulations similar to the ones leading to (\ref{transfoleadingto}) in Appendix A, we recover (\ref{firstorderinfluencefunctional1}). Therefore, reorganizing the perturbative expansion, we find that (\ref{premierTermeQuad}) gives rise to a correction to the effective action identical to the one calculated in (\ref{classicalfirstorderaction}), with $\xi_{1,2}^{\phi,\psi}$ interpreted as random classical Gaussian noises with variance given by the probability distribution (\ref{noiseprobdistribution}). The only difference here is that the mode functions used to calculate the $ {\bf M}_\psi^{i,j}(k\tau, k\tau')$ matrices given by (\ref{theMmatrices}) must be solutions to the equations of motion including interactions, $\Lambda^{(2)}_{\phi}(k)\phi_{{\bf k}}=0$, $\Lambda^{(2)}_{\psi}(k)\psi_{{\bf k}}=0$.

\begin{widetext} 
Then, equation (\ref{deuxiemetermQuad}) leads to a contribution at one loop given by the diagrams:

\begin{figure}[h]
\begin{center}
\includegraphics[width=12cm]{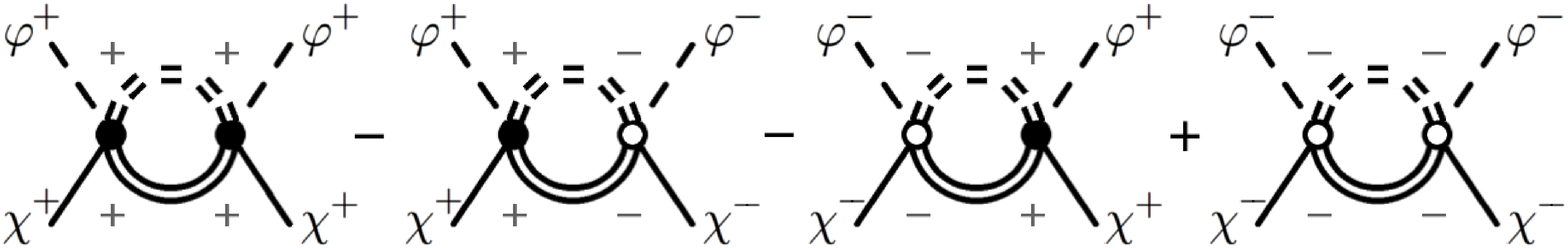}
\label{fig:oneloop}
\end{center}
\end{figure}

Evaluating these graphs, we obtain:
\bea
	-\frac{2g^4}{\hbar^2}\int d^4xd^4x' a^3(t)\left[  \varphi^+(x)\chi^+(x),~  \varphi^-(x)\chi^-(x)\right] \left( \begin{array}{c c} T \langle\phi^+_x \phi^+_{x'}\rangle T\langle\psi^+_x \psi^+_{x'}\rangle& -\langle\phi^-_{x'} \phi^+_{x}\rangle\langle\psi^-_{x'} \psi^+_x\rangle \\ -\langle\phi^-_{x} \phi^+_{x'}\rangle\langle\psi^-_{x} \psi^+_{x'}\rangle & \bar{T} \langle\phi^+_x \phi^+_{x'}\rangle \bar{T}\langle\psi^+_x \psi^+_{x'}\rangle \end{array}\right)\left[\begin{array}{c} \varphi^+(x')\chi^+(x')\\ \varphi^-(x')\chi^-(x') \end{array}\right] a^3(t')\, .\nonumber\\
\eea
Rotating to the Keldysh basis, we obtain, up to second order in the quantum fields (terms of third and fourth order in $\chi_q$ and $\varphi_q$ will also receive contributions from two and higher loops, and give higher contributions to the one-loop Langevin equation, we therefore neglect them in the following), the following contribution to the influence action:
\bea
\label{1-loopcontribbig}
	\delta S^{1-loop}_{\uIA}=\qquad\qquad\qquad\qquad\qquad\qquad\qquad\qquad\qquad\qquad\qquad\qquad\qquad\qquad\qquad\qquad\qquad\qquad\qquad\qquad\qquad\qquad \\
	2 g^4 \frac{i}{\hbar}\int d^4x d^4x' a^3(t) \left( \varphi_c (x)\chi_q(x)+\varphi_q (x)\chi_c(x) \right)\mathrm{Re} \left[ \Omega_{\phi}(x,x')\Omega_{\psi}(x,x') \right]   \left( \varphi_c (x')\chi_q(x')+\varphi_q (x')\chi_c(x') \right) a^3(t')
 \nonumber\\
	-\frac{8 g^4}{\hbar}\int d^4x d^4x' a^3(t)\varphi_c (x)\chi_c(x)\theta (t-t') \mathrm{Im} \left[ \Omega_{\phi}(x,x')\Omega_{\psi}(x,x') \right]   \left( \varphi_c (x')\chi_q(x')+\varphi_q (x')\chi_c(x') \right)a^3(t')\, .\nonumber
\eea

As before, we obtain both a real and an imaginary contribution to the influence action, which we interpret as a a noise and dissipation term, respectively. We again see a manifestation of the fluctuation-dissipation theorem. As before, the dissipation term will be negligible in the context of slow-roll inflation where most of the dissipative effects come from the Hubble friction term in the equations of motion. Also, again as before, in order to properly exhibit the noise as so, we introduce the classical random noise terms $\xi^\phi_3$ and $\xi^\psi_3$ defined by the probability distribution:
\be
\label{noiseprobdistribution1Loop}
	\mathcal{P}\left[\xi^\phi_3, \xi^\psi_3 \right]= \exp\left[ -\frac{1}{2}\int d^4xd^4x' [\xi^\phi_3 (x), \xi^\psi_3(x)] {\bf B}^{-1}(x,x')\left[\begin{array}{c} \xi^\phi_3(x')\\  \xi^\psi_3(x') \end{array}\right] \right]\, ,
\ee
where, if we assume that the external momenta of the coarse-grained fields are small compared to the internal momenta running in the loop, the {\bf B} matrix can be simplified as follow:
\be
\label{the1loopmatrix}
	{\bf B}(x,x')=\int^{a\Lambda} \frac{d k}{2\pi^2} k^2   \left[W_H\left(k\tau\right)W_H \left(k\tau'\right)\right]^2 \mathrm{Re}\left[ {\bf N}^{i,j}(k\tau, k\tau')\right] \delta^{3}({\bf x} - {\bf x}') + c.t.\, ,
\ee
with
\bea
 {\bf N}^{i,j}(k\tau, k\tau')=\phi_{{\bf k}}(t)\phi^*_{{\bf k}}(t')\psi_{{\bf -k}}(t) \psi^*_{{\bf -k}}(t') \left(\begin{array}{cc} 1  &  1 \\  1& 1  \end{array}\right)\, .
\eea
\end{widetext}
Although IR finite thanks to the presence of the coarse-graining window functions, the integrals in (\ref{the1loopmatrix}) will be UV divergent once we plug in an explicit mode function with Bunch-Davis initial conditions (as opposed to the integral in (\ref{themonstercorrelationmatrix}), which was both UV and IR finite due to the presence of derivative of window functions, effectively mimicking the effect of delta functions). However these UV divergences are easily dealt with using the standard procedure by introducing a physical UV cutoff $\Lambda$ and a local counter term (which we denote $c.t.$ in the above), as in \cite{Senatore2009}. This translates into a UV cutoff in the comoving momentum $k$ at $a\Lambda$, where the scale factor $a$ is evaluated at the smallest of the two times $t$, $t'$. The latter must then also be regulated with the same physical cutoff, which is most easily done by rewriting the time integrals as:
\bea
	\frac{1}{2}\int^{t_0} dt dt'=\int^{t_0} dt \int^{t}dt'\rightarrow \int^{t_0} dt \int^{t-\frac{1}{\Lambda}}dt'
\eea

The local counter term will then cancel any power law divergence, while replacing logarithmic divergences by terms proportional to $\ln \left(\frac{H}{\mu}\right)+C$, where $\mu$ is the renormalization scale, and $C$ is a constant.

The contribution to the influence action coming from the term containing the real part of the kernel $\Omega_\phi(x,x')\Omega_\psi(x,x')$ in (\ref{1-loopcontribbig}) then becomes:
\bea
\label{correctionforoneinterractingloop}
	\delta S^{1-loop}_{\uIA}=\qquad \qquad\qquad \qquad\qquad \qquad\qquad \qquad\qquad \\
	 2g^2 \int d^4 xa^3(t) \left[\xi^\phi_3(x), ~ \xi^\psi_3(x)\right] \left[ \begin{array}{c} \varphi_c(x)\chi_q(x) \\ \chi_c(x)\varphi_q(x) \end{array} \right]\, .\nonumber
\eea

\subsection{Contributions of the $\ln \Det$ terms at one loop }

To write down the one-loop Langevin equations, we only have left to deal with the two $\ln \Det (\tilde \Lambda_{\phi, \psi}^{(2)} )^{-1/2}$ terms in (\ref{unperturbedinfluencefunctionalTEXT}), as they contain resumed powers of the system fields which need to be differentiated in order to obtain the right equations of motion.  Calculating its influence to first order as done in Appendix C, we obtain a non-local mass renormalization term, a new noise term, as well as a new dissipation term. Once rotated to the Keldysh basis, we obtain for the $\varphi$ and $\chi$ fields, keeping only corrections up to one-loop order:
\begin{widetext}
%

\bea
\label{detlogpsi}
	-\frac{1}{2}\ln \Det (\tilde \Lambda_{\phi}^{(2)} )&=& \frac{i}{\hbar}\left\lbrace -\frac{g^2}{2} \int d^4x a^3(t)  \chi_c \chi_q  F_{\phi\phi}(x,x)\right. \qquad\qquad \qquad \qquad\qquad \qquad \qquad\qquad \qquad \qquad\qquad \qquad \qquad\qquad  \\
		&&+ g^4 \frac{i}{\hbar}\int d^4x d^4x' a^3(t) \chi_c (x)\chi_q(x)\mathrm{Re} \left[  F_{\phi\phi, \phi\phi}(x, x') \right]   \chi_c (x')\chi_q(x') a^3(t') \nonumber\\
	&&\left.-\frac{2 g^4}{\hbar}\int d^4x d^4x' a^3(t)\chi_c (x)\chi_c(x)\theta (t-t') \mathrm{Im} \left[   F_{\phi\phi, \phi\phi}(x, x') \right]    \chi_c (x')\chi_q(x')a^3(t')\right\rbrace\nonumber \, ,
\eea

\bea
\label{detlogphi}
	-\frac{1}{2}\ln \Det (\tilde \Lambda_{\psi}^{(2)} )= \frac{i}{\hbar}\left\lbrace-\int d^4x a^3(t) \left( \frac{\lambda}{4} \chi_c \chi_q +  \frac{g^2}{2}\varphi_c \varphi_q\right)\right.F_{\psi\psi}(x,x)\qquad \qquad \qquad\qquad \qquad \qquad\qquad \qquad \qquad\qquad \qquad \\
	+  \frac{i}{\hbar}\int d^4x d^4x' a^3(t)\left[g^2\varphi_c (x)\varphi_q(x),  ~ \frac{\lambda}{2}\chi_c (x)\chi_q(x) \right]\mathrm{Re} \left[ {\bf F}_{\psi\psi, \psi\psi}(x, x')\right]   \left[\begin{array}{c}g^2\varphi_c (x')\varphi_q(x') \\  \frac{\lambda}{2}\chi_c (x')\chi_q(x') \end{array}\right] a^3(t')
 \nonumber\\
	\left.-\frac{2}{\hbar}\int d^4x d^4x' a^3(t)\theta (t-t') \left[ g^2\varphi_c (x)\varphi_c(x),~\frac{\lambda}{2}\chi_c (x)\chi_c(x)\right] \mathrm{Im} \left[ {\bf F}_{\psi\psi, \psi\psi}(x, x') \right]    \left[\begin{array}{c}g^2\varphi_c (x')\varphi_q(x') \\  \frac{\lambda}{2}\chi_c (x')\chi_q(x') \end{array}\right] a^3(t')\right\rbrace\nonumber\, ,
	\eea
where the different $F$'s above are related to different powers of the propagators of the free, massive, bath theory. In our case, for the purpose of giving an example, we can write them using the massless mode function for the inflaton field since we assume it to be light, and the full, massive mode function for the spectator field. Their regulated expression can then be explicitly written as: 
\bea
F_{\psi\psi}(x,x)&= &\int^{a\Lambda} \frac{dk}{2\pi^2} k^2\left(W_H^t\right)^2\frac{ H^2\pi}{4}  (-\tau)^3 \left|H_\nu (-k\tau)\right|^2 + c.t.\, ,\\
F_{\phi\phi}(x,x)&=& \int^{a\Lambda} \frac{dk}{2\pi^2} \left(W_H^t\right)^2 \frac{ H^2}{2}   \left(\frac{1}{k} + k\tau^2\right) + c.t. \, ,
\eea
\bea
\label{thedetlog1loopmatrices}
	F_{\phi\phi, \phi\phi}(x, x')=\int^{a \Lambda} \frac{d k}{2\pi^2} k^2   \left[W_H\left(k\tau\right)W_H \left(k\tau'\right)\right]^2 \mathrm{Re}\left[ O_\phi(k\tau, k\tau')\right] \delta^{3}({\bf x} - {\bf x}') +c.t.\, ,\\
	{\bf F}_{\psi\psi, \psi\psi}(x, x')=\int^{a \Lambda} \frac{d k}{2\pi^2} k^2   \left[W_H\left(k\tau\right)W_H \left(k\tau'\right)\right]^2 \mathrm{Re}\left[ {\bf O}_\psi^{i,j}(k\tau, k\tau')\right] \delta^{3}({\bf x} - {\bf x}') +c.t.\, ,
\eea
where `$c.t.$' denote the contribution of a counter term, and where:
\bea
  O_\phi(k\tau, k\tau')=\frac{ H^4}{4k^6}  \left[1+k^2\tau\tau' +ik(\tau-\tau')\right]^2e^{-2ik(\tau-\tau')}\, ,\\
 {\bf O}_\psi^{i,j}(k\tau, k\tau')=\frac{ H^4\pi^2}{16}  (\tau\tau')^{3}\left[H_\nu (-k\tau)H^*_\nu (-k\tau')\right]^2\left(\begin{array}{cc} 1  &  1 \\  1& 1  \end{array}\right)\, .
\eea
Here, we use $\nu$ to denote the usual mass-correction factor in the Hankel functions: $ \nu =\sqrt{ 9/4-m_\psi^2/H^2}$. 
\bigskip

Introducing the extra classical gaussian noise terms $\xi^\phi_4$, $\xi^\psi_4$ and $\xi_5$, defined by the probability distributions:
\be
	\mathcal{P}\left[\xi^\phi_4, \xi^\psi_4 \right]= \exp\left\lbrace -\frac{1}{2}\int d^4xd^4x' [\xi^\phi_4 (x), \xi^\psi_4(x)] {\bf F}_{\psi\psi, \psi\psi}^{-1}(x,x')\left[\begin{array}{c} \xi^\phi_4(x')\\  \xi^\psi_4(x') \end{array}\right] \right\rbrace\, ,\nonumber \\
\ee
\be
	\mathcal{P}\left[\xi_5 \right]= \exp\left[ -\frac{1}{2}\int d^4xd^4x' \xi_5 (x) F_{\phi\phi, \phi\phi}^{-1}(x,x') \xi_5(x')\right] \, ,
\ee
and neglecting the dissipation terms in (\ref{detlogphi}) and (\ref{detlogpsi}), we can then rewrite the two  $\ln \Det (\tilde \Lambda_{\phi, \psi}^{(2)} )^{-1/2}$ terms contribution to the influence action up to one loop as:
\bea
	\delta S^{\ln \Det}_{\uIA}= \qquad \qquad \qquad \qquad\qquad \qquad \qquad\qquad  \qquad\qquad \qquad\qquad \qquad \qquad\qquad \qquad \qquad\qquad \qquad \qquad\qquad \qquad\\
	\int d^4x a^3(t) \left\lbrace-\left( \frac{\lambda}{2} \chi_c \chi_q +  g^2\varphi_c \varphi_q\right)F_{\psi\psi}(x,x)-g^2 \chi_c \chi_q  F_{\phi\phi}(x,x) +2\left[\xi^\phi_4,  ~ \xi^{\psi}_4 \right]  \left[\begin{array}{c}g^2\varphi_c (x')\varphi_q(x') \\  \frac{\lambda}{2}\chi_c (x')\chi_q(x') \end{array}\right]\nonumber 
	+2 g^2 \xi_5 \chi_c (x)\chi_q(x)\right\rbrace
\eea
Therefore, bringing all the pieces together, we can collect the above contribution to the influence action together with what we calculated from the one-loop interaction in (\ref{correctionforoneinterractingloop}) and add all these interacting one-loop corrections to the free, leading order result we have calculated in (\ref{classicalfirstorderaction}). Since we are, as before, interested in exploring the classical limit of the effective action, we once again proceed to the field rescaling $\varphi_q$, $\chi_q\rightarrow\hbar \varphi_q$, $\hbar\chi_q$. We again want to perform a first order expansion of the influence action in $\hbar \varphi_q$, $\hbar \chi_q$, which corresponds to the $\hbar$-independent, or classical, sector of the action:

\bea 
 	\int^{\varphi^+}_{\varphi^\pm_i}\mathcal{D}\varphi^\pm\int^{\chi^+}_{\chi^\pm_i}\mathcal{D} \chi^\pm  \exp\left[\frac{i}{\hbar} S^{(1)}_{\ueff}\right]=\int \mathcal{D}\varphi^{q,c}\mathcal{D} \chi^{q,c} \int\mathcal{D}\xi^{\phi,\psi}_{1,2,3,4} \mathcal{D}\xi_5 \mathcal{P}\left[\xi^{\phi,\psi}_{1}, \xi^{\phi,\psi}_2 \right] \mathcal{P}\left[\xi^{\phi}_{3,4}, \xi^{\psi}_{3,4} \right] \mathcal{P}\left[\xi_5 \right]\qquad \qquad\qquad\\
	\times \exp\left( i\int d^4x a^3(t)\left\lbrace\varphi_q\left[ \left(\Box-m_\Phi^2\right) \varphi_c - {V}_{\mathrm{pert},\Phi}\left(\varphi_c, \chi_c\right) \right] +\chi_q\left[ \left(\Box-m_\Psi^2\right)\chi_c -{V}_{\mathrm{pert},\Psi}\left(\varphi_c, \chi_c\right) \right]\nonumber \right. \right.\\
	  +\varphi_q\left[ p_\phi(t)\xi^\phi_1+\xi^\phi_2\right]-\dot{\varphi}_q\xi^\phi_1+\chi_q\left[ p_\psi(t)\xi^\psi_1+\xi^\psi_2\right]-\dot{\chi}_q\xi^\psi_1
	 +2g^2 \left[\xi^\phi_3(x), ~ \xi^\psi_3(x)\right] \left[ \begin{array}{c} \varphi_c(x)\chi_q(x) \\ \chi_c(x)\varphi_q(x) \end{array} \right] \nonumber\\
	 \left.\left. -\left( \frac{\lambda}{2} \chi_c \chi_q +  g^2\varphi_c \varphi_q\right)F_{\psi\psi}(x,x)-g^2 \chi_c \chi_q  F_{\phi\phi}(x,x) +2\left[\xi^\phi_4,  ~ \xi^{\psi}_4 \right]  \left[\begin{array}{c}g^2\varphi_c (x')\varphi_q(x') \\  \frac{\lambda}{2}\chi_c (x')\chi_q(x') \end{array}\right]\nonumber 
	+2 g^2 \xi_5 \chi_c (x)\chi_q(x)  \right\rbrace\right) \,.
\label{classicalonelooporderaction}
\eea
The equations of motion for the classical fields $\varphi_c$ and $\chi_c$ are then obtained via the saddle-point conditions (\ref{langevinequationssaddlepointcondition}):
\bea
\label{stochasticphi1loopfromCTP}
	(-\Box+m_\Phi^2)\varphi_c +\tilde{V}_{,\Phi}(\varphi_c, \chi_c)&=&p_\phi(t)\xi^\phi_1+\xi^\phi_2+\dot{\xi}_1^\phi+3H\xi^\phi_1 
	+2g^2\xi^\phi_3(x) \varphi_c(x) +2g^2\xi^\phi_4 \varphi_c (x) \nonumber \\ 
	&&- \frac{\lambda}{2} \chi_c F_{\psi\psi}(x,x)\, ,\\
	(-\Box +m^2_\Psi)\chi_c+ \tilde{V}_{,\Psi}(\varphi_c, \chi_c)&= &p_\psi(t)\xi^\psi_1+\xi^\psi_2+\dot{\xi}_1^\psi+3H\xi^\psi_1
	+2g^2 \xi^\psi_3(x)\chi_c(x)\nonumber \\
	&&- g^2\varphi_c F_{\psi\psi}(x,x) -g^2 \chi_c  F_{\phi\phi}(x,x) +\lambda\xi^{\psi}_4\chi_c (x) +2 g^2 \xi_5 \chi_c (x),
\label{stochasticpsi1loopfromCTP}
\eea
where the mode functions used to find the variance of the various noise terms in the above are the one solving the equations of motion:
\bea
\left[\Box-m_\Phi^2-g^2\left(\chi^\pm\right)^2\right]\phi_{{\bf k}}(t)&=&0\,,\nonumber\\
\left[\Box-m_\Psi^2-g^2\left(\varphi^\pm\right)^2-\frac{\lambda}{2}\left(\chi^\pm \right)^2\right]\psi_{{\bf k}}(t)&=&0.
\eea

\end{widetext}
This is the main result of the paper. We clearly see that the extra terms derived at one loop are actually introducing back-reaction effects in
the sense that they couple two bath fields with a (classical) system degree of freedom as a source in the classical fields effective equation of motion. When interested in the problem of back-reaction, the terms appearing in the above Langevin equations will be the leading ones in powers of the coupling constants of the theory embodying those mode coupling effects.

Linking back to our heuristic derivation from section II, these are the leading contributions to the extra terms in the Langevin equations we found in (\ref{stochastichybridfulleomphi}) and (\ref{stochastichybridfulleompsi}). More specifically, these are the precise terms which were second order in the quantum fields and which we wrote explicitly. This means that these second-order mode coupling terms in the quantum fields can in fact be rewritten as a sum of three different noise terms for the $\chi$ field and two different noise terms for the $\varphi$ field (closely linked to the fact that there are three really distinct diagrams at second order in the couplings contributing to the $\chi$ effective potential, while there are two for the $\varphi$ effective potential).

It is interesting to see that in the heuristic case, the method used could not capture the contributions to the equations of motions coming from the mass-renormalization terms in the effective potential of the system fields (which are the terms on the RHS of the above without noise terms), as well as the dispersion terms, which ended up being negligible in the slow-roll approximation and which we neglected in the above. Note that this was also the case when we evaluated the free influence functional in section III, where we obtain a real term for each system field giving rise to dispersion but decided to neglect it because their contribution was again negligible under the slow-roll assumption. The heuristic approach did not allow us to capture such effects, even at the free-action level. Indeed, had we kept them in the equations of motion, the dissipation terms we obtained in (\ref{firstorderinfluencefunctional1}) would have given rise to two friction terms (one per field) with sign opposite to the usual Hubble friction term. This anti-dissipation is therefore of a very different nature, and, just like the noise terms, they can be interpreted as coming from the constantly incoming modes through the window function into the coarse-grained theory. The fact that usual projection methods from the equations of motions miss this effect interestingly seems to suggest that dissipation arising from the fluctuation-dissipation theorem is a quantum effect (even though in the current case it turns out to be a subdominant one) which semi-classical methods, at least in their usual form, cannot capture.

As a final note, we would like to emphasize that throughout this paper, mainly for the sake of clarity, we have been working with fluctuating fields on a fixed classical background metric. In order to be fully consistent, one should explicitly calculate every noise term defined in terms of the slow-roll parameters, and keep gravitational corrections consistently up to whatever order we obtain. A priori, this might require expanding the equations of motion for the mode functions up to second order in metric perturbations, which would make things more complicated. 

\section{Conclusion}
\label{sec:conclusion}
In this paper, we have presented a heuristic but precise derivation of the Langevin equations of motion for stochastic inflation. We then have presented an alternative, more rigorous derivation from first principles of the formalism of stochastic inflation and pointed out the link between the presence of noise and the processes of decoherence and of classicalization. We then developed a self-consistent recursive method for obtaining a solution at consistent order in fields perturbations and in slow-roll, motivated by the case-study of multi-field hybrid inflation which we explore in a follow-up paper, where it has been suspected for many years that quantum effects and back-reaction can have a significant effects. 

In line with this, we developed a better understanding of the loop expansion both in the bath fields and in the system-bath coupling within the CTP framework. In particular, we found how the term representing mode coupling found in the heuristic derivation emerges from the perturbative expansion.

In the second paper of this series \cite{secondpaper}, we will apply our new recursive method to the specific example of two-field hybrid inflation potentials, as an example of how this method can be implemented in a generic multi-field context. We calculate the modified predictions emerging from a consistently-implemented non-perturbative method for cosmological observables such as the CMB power spectrum. Most interestingly, we identifiy regimes of hybrid inflation where stochastic effects dominate over regular perturbative corrections. This result has potentially important implications for accurately constraining such small-field models of inflation.

In future work, we wish to apply our method including mode coupling effects during the waterfall of hybrid-type models. This is of particular interest, since stochastic effects have been shown to completely dominate the classical evolution of the background fields trajectory in this phase \cite{Martin2011}. Furthermore, there is plenty of scope for applying the method developed in the present paper to phenomenological models of inflation, an avenue which we plan to explore in future work.

\acknowledgements

The author would like to thank her collaborators, Vincent Vennin and Robert Brandenberger, who constantly helped with discussions and advices throughout the elaboration of this work, and carefully proofread the manuscript. The author is also very much grateful to Anne Davis, for her support, advices, help and many useful discussions. The author would also like to thank Jeremy Sakstein, Guy Moore, Hiro Funakoshi, Daniel Baumann and Valentin Assassi for helpful discussions.

This research is supported in part by a GPS-3D scholarship from NSERC of Canada, and by an M.T. Meyers Scholarship from Girton College, at the University of Cambridge.

\newpage


\begin{widetext}
\renewcommand{\theequation}{A \arabic{equation}}
\renewcommand{\thefigure}{A \arabic{figure}}
\setcounter{equation}{0}  
\setcounter{figure}{0}
\setcounter{subsubsection}{0}
\section*{APPENDIX A}  
\Large\begin{center}
{\scshape Path integral over the bath degrees of freedom : \\Leading order}
\\
\end{center}
\normalsize

Using the bath-system splitting defined in sections \ref{subsec:heuristicderivation} and \ref{subsec:SKderiv}, as well as the Fourier modes decomposition of the bath degrees of freedom defined in (\ref{quantumphipsiexpanded}), in the following our goal will be to perform an integration of the action over the bath degrees of freedom (the short wavelength fields), $\phi_>$ and $\psi_>$, assuming the full fields $\Phi$ and $\Psi$ are massive but have no interacting potential, in order to obtain an effective action $S_{\ueff}$ for the system of long-wavelength modes, and finally take the classical limit.

In other words, we want to look at how to perform the path integral for the leading terms including bath fields, that is, the bilinear piece of the action. As we will see, this term will give rise to the noise terms we encountered in (\ref{stochastichybridfulleomphi}) and (\ref{stochastichybridfulleompsi}). Using the notation we introduced in (\ref{inflienceactionlong}) - (\ref{integrationkernelsforscalarfields}) of section \ref{subsec:SKderiv}, we want to calculate:
\bea
	F[\tilde{\varphi}, \tilde{\chi}]&=&\int_{-\infty}^{\infty}d\phi_{>_f}^+ d\psi_{>_f}^+ \int \mathcal{D} \phi_{>}^{\pm}\int\mathcal{D}\psi_{>}^{\pm} e^{\frac{i}{\hbar} \left[  \tilde{S}^+_{\ubath}-\tilde{S}_{\ubath}^- \right]}\nonumber\\
	&=&\int_{-\infty}^{\infty}d\phi_{>_f}^+ d\psi_{>_f}^+ 
	\int \mathcal{D} \phi_{>}^{\pm}  e^{\frac{i}{\hbar}\int d^4x\left[ \frac{1}{2} \tilde{\phi} ^T_>\tilde{\Lambda}_\phi \tilde{\phi}_>  + \tilde{\varphi} ^T\tilde{\Lambda}_\phi \tilde{\phi}_>  \right]}\int\mathcal{D}\psi_{>}^{\pm} e^{\frac{i}{\hbar}\int d^4x\left[ \frac{1}{2}\tilde{\psi} ^T_>\tilde{\Lambda}_\psi\tilde{ \psi}_>+ \tilde{\chi} ^T\tilde{ \Lambda}_\psi \tilde{ \psi}_>  \right]} \nonumber \, ,\\
\eea
where we are still using the matrix notation introduced in (\ref{matrixnotationshort}). 
{Since $\tilde{\Lambda}_\phi$ and $\tilde{\Lambda}_\phi$ are symmetrical operators with positive eigenvalues,} we can then complete the square in the exponential and perform the Gaussian path integral over the bath fields:
\bea
	F[\tilde{\varphi}, \tilde{\chi}]&=&N\exp\left[ -\frac{1}{2}\frac{i}{\hbar}\left( \int d^4 x d^4 x' \tilde{\varphi}^T(x)\vec{\tilde{\Lambda}}_\phi (\tilde{\Lambda}_\phi)^{-1}  \cev{\tilde{\Lambda}}_\phi\tilde{\varphi}(x') + \int d^4 y d^4 y' \tilde{\chi}^T(y)\vec{\tilde{\Lambda}}_\psi (\tilde{\Lambda}_\psi)^{-1}  \cev{\tilde{\Lambda}}_\psi \tilde{\chi}(y')\right)\right],
\eea
where $N$ is a normalization factor in which we absorbed factors of square root of determinant of $\tilde{\Lambda}$, and the arrows over $\tilde{\Lambda}_\phi$ and $\tilde{\Lambda}_\psi$ indicate the direction in which the time derivatives they contain are acting. Also, here, $(\tilde{\Lambda}_\phi)^{-1}$ and $(\tilde{\Lambda}_\psi)^{-1}$ are Green's functions on the contour for the bath fields, $\phi_>$ and $\psi_>$ respectively, in the CTP method:
\be
\label{firstdefofprops}
	i\hbar(\tilde{\Lambda}_\phi)^{-1}=\left(\begin{array}{cc} \left\langle T\hat\phi^+_>(x)\hat\phi^+_>(x') \right\rangle &  \left\langle \hat\phi^-_>(x')\hat\phi^+_>(x) \right\rangle  \\   \left\langle \hat\phi^-_>(x)\hat\phi^+_>(x') \right\rangle &  \left\langle \bar{T}\hat\phi^-_>(x)\hat\phi^-_>(x') \right\rangle \end{array}\right)\, , \quad
	i\hbar(\tilde{\Lambda}_\psi)^{-1}=\left(\begin{array}{cc} \left\langle T\hat\psi^+_>(y)\hat\psi^+_>(y') \right\rangle & \left\langle\hat \psi^-_>(y')\hat\psi^+_>(y) \right\rangle  \\  \left\langle \hat\psi^-_>(y)\hat\psi^+_>(y') \right\rangle &  \left\langle \bar{T}\hat\phi^-_>(y)\hat\phi^-_>(y') \right\rangle \end{array}\right)\, ,
\ee  
The entries of these propagators can easily be interpreted intuitively. The upper-left entry in each propagator represents the amplitude for propagation between point $x$ and $x'$ on the ``$+$"-part (or forward part) of the contour. The lower-right entry represent the amplitude for propagation if both $x$ and $x'$ are on the ``$-$" or backward part of the contour. The off-diagonal entries represent the amplitude for a particle to propagate from an ``$x$" on the forward part of the contour to and ``$x'$" located on the backward part, and reversely. 

Writing the fields in Fourier space, and recalling that the bath fields are defined as in (\ref{quantumphipsiexpanded}), we can pull out of the propagator one factor of the window function on each side of each propagator:
\bea
	 F[\tilde{\varphi}, \tilde{\chi}]&=&N\exp\left[ -\frac{i}{2\hbar} \int dt dt' \int \frac{d^3{\bf k}}{(2\pi)^3} \frac{d^3{\bf q}}{(2\pi)^3} \tilde{\varphi}_{-{\bf k}}^T(t)\vec{\tilde{\Lambda}}_\phi(t) W_H\left(\frac{k}{\epsilon aH} \right) \right. \nonumber \\
	&&\qquad\times \left. \frac{-i}{\hbar}\left(\begin{array}{cc} \left\langle T\hat\phi_{\bf k}(t)\hat\phi_{\bf q}(t') \right\rangle &  \left\langle \hat\phi_{\bf k}(t')\hat\phi_{\bf q}(t) \right\rangle  \\   \left\langle \hat\phi_{\bf k}(t)\hat\phi_{\bf q}(t') \right\rangle &  \left\langle \bar{T}\hat\phi_{\bf k}(t)\hat\phi_{\bf q}(t') \right\rangle \end{array}\right)  W_H\left(\frac{q}{\epsilon aH}\right) \cev{\tilde{\Lambda}}_\phi(t') \tilde{\varphi}_{-{\bf q}}(t')\right] \nonumber \\
	&&\qquad \times\, \exp\left[  \varphi \leftrightarrow \chi,\,\phi \leftrightarrow \psi \right], 
\eea
where in the last line, we mean repeat the previous exponential term, but with all $\varphi$ and $\phi$ fields replaced by $\chi$ and $\psi$ fields.
Now, recall that the system fields $\varphi$ and $\chi$ are defined to be non-zero precisely on the long wavelengths where the window function $W_H$ was chosen to vanish. The orthogonality (in the $W\rightarrow \theta$ limit) of $\varphi_k$ and $\chi_k$ with $W_H$ means that the only non-zero terms in the above will be the ones where at least one time derivative is acting on all $W_H$'s.

\bea
	&=&N\exp\Bigg[  \frac{1}{2}\left(\frac{i}{\hbar}\right)^2  \int dt dt' \int \frac{d^3{\bf k}}{(2\pi)^3} \frac{d^3{\bf q}}{(2\pi)^3}a^3(t) \tilde{\varphi}_{-{\bf k}}^T(t) 
\vec{P}_t\sigma_3 \left(\begin{array}{cc} \left\langle T\hat\phi_{\bf k}(t)\hat\phi_{\bf q}(t') \right\rangle &  \left\langle \hat\phi_{\bf k}(t')\hat\phi_{\bf q}(t) \right\rangle  \\   \left\langle \hat\phi_{\bf k}(t)\hat\phi_{\bf q}(t') \right\rangle &  \left\langle \bar{T}\hat\phi_{\bf k}(t)\hat\phi_{\bf q}(t') \right\rangle \end{array}\right)\sigma_3 \cev{P}_{t'} a^3(t')\tilde{\varphi}_{-{\bf q}}(t')\Bigg] \nonumber \\
&&\qquad	\times \exp\left[\varphi \leftrightarrow \chi,\, \phi \leftrightarrow \psi \right] \label{eq:F1}\\
	&=&N\exp\left[ \frac{1}{2}\left(\frac{i}{\hbar}\right)^2 \int dt dt' \int \frac{d^3{\bf k}}{(2\pi)^3} a^3(t)\tilde{\varphi}_{-{\bf k}}^T(t)  \vec{P}_t\right.\times \nonumber \\
	  &&\qquad \left.  \left(\begin{array}{cc} \theta(t-t')\phi_{\bf k}(t)\phi^*_{\bf k}(t')+\theta(t'-t)\phi_{\bf k}(t')\phi^*_{\bf k}(t) & -\phi_{\bf k}(t')\phi^*_{\bf k}(t) \\  - \phi_{\bf k}(t)\phi^*_{\bf k}(t') &  \theta(t'-t)\phi_{\bf k}(t)\phi^*_{\bf k}(t')+\theta(t-t')\phi_{\bf k}(t')\phi^*_{\bf k}(t)  \end{array}\right) \cev{P}_{t'} a^3(t') \tilde{\varphi}_{{\bf k}}(t')\right]\nonumber \\
	&&\qquad \times \exp\left[ \varphi \leftrightarrow \chi,\, \phi \leftrightarrow \psi \right]\label{eq:F2}\\
	&=&N\exp\left[\frac{1}{2}\left(\frac{i}{\hbar}\right)^2 \int d^4x d^4x' a^3(t)\tilde{\varphi}^T(x)  \vec{P}_t\int \frac{d^3{\bf k}}{(2\pi)^3} \right.\times \nonumber \\
	  &&\qquad\left(\begin{array}{c} \theta(t-t')\phi_{\bf k}(t)e^{-i{\bf k}\cdot {\bf x}}\phi^*_{\bf k}(t')e^{i{\bf k}\cdot {\bf x'}}+\theta(t'-t)\phi_{\bf k}(t')e^{i{\bf k}\cdot {\bf x'}}\phi^*_{\bf k}(t)e^{-i{\bf k}\cdot {\bf x}} \\    - \phi_{\bf k}(t)e^{-i{\bf k}\cdot {\bf x}}\phi^*_{\bf k}(t')e^{i{\bf k}\cdot {\bf x'}}\end{array}\right. \nonumber \\ 
	  && \qquad \quad\qquad \quad\qquad \quad\qquad  \left.  \left.\begin{array}{c} -\phi_{\bf k}(t')e^{i{\bf k}\cdot{\bf x'}}\phi^*_{\bf k}(t)e^{-i{\bf k}\cdot{\bf x}} \\  \theta(t'-t)\phi_{\bf k}(t)e^{-i{\bf k}\cdot{\bf x}}\phi^*_{\bf k}(t')e^{i{\bf k}\cdot{\bf x'}}+\theta(t-t')\phi_{\bf k}(t')e^{i{\bf k}\cdot{\bf x'}}\phi^*_{\bf k}(t)e^{-i{\bf k}\cdot{\bf x}}  \end{array}\right) \cev{P}_{t'} a^3(t') \tilde{\varphi}(x')\right]\nonumber \\
	&&\qquad\times \exp\left[ \varphi \leftrightarrow \chi,\, \phi \leftrightarrow \psi \right]\, .\label{eq:F3}
\eea
where from  (\ref{eq:F1}) to (\ref{eq:F2}) we have used the operator $\vec{P}_t$ defined in (\ref{ptoperator}) as well as the expansion of the quantum fields in terms of creation and annihilation operators in order to evaluate the expectation values, $\hat\phi_{{\bf k}}=\phi_{{\bf k}}\hat a_{{\bf k}}+\phi^*_{{\bf k}}\hat a^\dagger_{{\bf k}}$ and similarly for $\hat\psi_{{\bf k}}$. Also from (\ref{eq:F2}) to (\ref{eq:F3}), we have re-expanded the fields $\tilde{\varphi}$ back in real space.

The explicit expressions for the propagators above demonstrate that the upper and lower contours are not independent. In fact, the Green's functions written above contain a certain degree of redundancy. For example, we have that the trace of the matrix above equals the sum of the off-diagonal terms. Therefore, defining the linear transformation:
\be
	U=\left( \begin{array}{cc} 1/2& 1/2 \\ 1 & -1 \end{array}\right)\, ,
\ee
we obtain an equivalent representation of the bloc propagator we were working with. In this new basis, called the Keldysh representation, the system fields take the form we used in (\ref{transfotoclassquant}) to define the quantum and classical fields:
\be
	\tilde{\varphi}'=U\tilde{\varphi} =\left( \begin{array}{c} \frac{\varphi^++\varphi^-}{2} \\ \varphi^+-\varphi^- \end{array}\right)\equiv\left(\begin{array}{c}\varphi_c \\ \varphi_q \end{array} \right)\, ; \qquad \tilde{\chi}'=U\tilde{\chi} =\left( \begin{array}{c} \frac{\chi^++\chi^-}{2} \\ \chi^+-\chi^- \end{array}\right)\equiv\left(\begin{array}{c}\chi_c \\ \chi_q \end{array} \right), 
\ee
where $\varphi_c$, $\chi_c$ and $\varphi_q$, $\chi_q$ define the classical and quantum fields, respectively \cite{Altland2010}. In terms of these new variables, and making use of the identities Re$\left[ ab^* \right]=(ab^*+a^*b)/2$ and Im$[ab^*]=(ab^*-a^*b)/2$ we obtain (\ref{firstorderinfluencefunctional1}) after a short calculation. We therefore find the real and imaginary parts of $\Pi_{\phi,\psi}(x, x')$ contributing to the influence functional, which we can interpret as dissipation and noise kernels coming from the bath degrees of freedom, respectively, as discussed in section \ref{subsec:SKderiv}. However, in order to be able to properly interpret the imaginary part of the influence functional as noise, we now proceed to a few aesthetic manipulations on the operator (\ref{Pioperator}).

First we rewrite $P_t$ in a way that it contains only first time derivatives of the window function $W_H$:
\bea
a^3(t)
	\left[P_t\phi_{{\bf k}}(t)\right]e^{-i{\bf k}\cdot{\bf x}} &=& \left\lbrace \partial_t\left[a^3(t)\dot{W}_H\right]+2a^3(t)\dot{W}_H\partial_t \right\rbrace\phi_{{\bf k}}(t)e^{-i{\bf k}\cdot{\bf x}}\nonumber \\
	&=&  \partial_t\left[a^3(t)\dot{W}_H\phi_{{\bf k}}(t)e^{-i{\bf k}\cdot{\bf x}}\right]+a^3(t)\dot{W}_H\partial_t \phi_{{\bf k}}(t)e^{-i{\bf k}\cdot{\bf x}}\nonumber\\
	&=& \partial_t\left[a^3(t)\dot{W}_H\phi_{{\bf k}}(t)e^{-i{\bf k}\cdot{\bf x}}\right]+a^3(t)\dot{W}_H \left\lbrace p_\phi(t)+ q_\phi \left[\frac{k}{a(t)H(t)}\right]\right\rbrace \phi_{{\bf k}}(t)e^{-i{\bf k}\cdot{\bf x}}\,.
	\label{thePfunctiontointbyparts}
\eea
In the last line, for the mode functions of the quantum field we have assumed the form $\dot{\phi}_{\bf k}=\left\lbrace p_\phi (t)+q_\phi \left[k/(aH)\right]\right\rbrace\phi_{\bf k}$ with some function of time $q$ which contains all the $k$-dependence through the combination $k/aH=-k\tau (1-\varepsilon)$ (to first order in slow-roll, where $\tau$ is, as before, the conformal time, and $\varepsilon\equiv-\dot{H}/H^2$ is the first slow-roll parameter) and some function of time $p$ with no $k$-dependence. This is the case whenever mode functions can be written under the form of a Hankel function of the first kind, which is what we will be assuming in the following.
For concreteness, in the case of single-field inflation, for a field with mass $m$ where we neglect metric perturbations, we find:
\be
\label{ladefdepphipsi}
	p(t)+q\left(-k\tau\right)=-H(t)\left[ \frac{3}{2}-\nu+\frac{\dot{H}}{H^2}\left( -\nu+\frac{1}{2}\right)+(-k\tau) \frac{H^{(1)}_{\nu-1}\left(-k\tau\right)}{H^{(1)}_\nu\left(-k\tau\right)}\left(1+\frac{\dot{H}}{H^2} \right) \right]\,,
\ee
where $\nu^2$ assumes the standard value of $\frac{9}{4}-\frac{m^2}{H^2}+3\varepsilon$ (to first order in slow-roll). In general, the exact functions $p$ and $q$ will be different for the two fields $\phi$ and $\psi$ we are considering here, and will also depend on the level of approximation we are making for solving the equations of motion of the mode functions. 

Once substituted back in (\ref{firstorderinfluencefunctional1}), the first term of (\ref{thePfunctiontointbyparts}) can be integrated by parts to yield (recall that, as explained in the main body of the text, the second term of (\ref{thePfunctiontointbyparts}) gives rise to both a mass-renormalization term and dissipation, which are negligible in the context of slow-roll inflation \cite{Morikawa1990, Matarrese2004}):
\bea
\label{transfoleadingto}
	S_{\uIA}&=& \frac{i}{2\hbar}\int d^4x d^4x'a^3(t)a^3(t')\int\frac{d^3{\bf k}}{(2\pi)^3} \left\lbrace \left[ p_\phi(t)+q_\phi \left(-k\tau\right)\right] \varphi_q(x) -\partial_{t} \varphi_q(x)\right\rbrace\nonumber \mathrm{Re}\left[\dot{W}(t)\phi_{{\bf k}}(t)\dot{W}(t')\phi^*_{{\bf k}}(t') e^{-i{\bf k}\cdot ({\bf x}-{\bf x}^\prime)}\right] \\ 
	&& \qquad\qquad \times\left\lbrace \left[p_\phi^*(t')+q^*_\phi \left(-k\tau' \right)\right] \varphi_q(x') -\partial_{t'} \varphi_q(x')\right\rbrace\nonumber\\
	&&+\left(\chi\leftrightarrow  \varphi \right)\,,
\eea
as argued in section \ref{subsec:heuristicderivation}, $W_H$ is constrained to have the form $W_H\left(\frac{k}{\epsilon aH}\right)$. Replacing $\dot{W}_H= -k/(a\epsilon)\left(1+\dot H/H^2 \right)\left( \partial_{k/(\epsilon aH)}W_H\right) \approx H k\tau \left( 1-\varepsilon \right)\partial_{k\tau}W_H$ (where the last equality holds to first order in slow-roll), we can perform the angular part of the integral over $d^3{\bf k}$ and rewrite the above expression more conveniently in matrix form:
\bea
	S_{\uIA}&=& \frac{i}{2\hbar}\int d^4x d^4x'a^3(t)a^3(t')\left[ \sigma^\phi_q(x),\, \varphi_q(x) \right] {\bf A}^{i,j}_\phi(x, x') \left[\begin{array}{c} \sigma^\phi_q(x')\\  \varphi_q(x')\end{array}\right]+\left(\chi\leftrightarrow  \varphi \right)\,,
\eea
where we have defined $\sigma^\phi_q(x)=\left[ p_\phi(t)-\partial_{t}\right] \varphi_q(x)$ and where the operators ${\bf A}^{i,j}_{\phi, \psi}(x, x')$ are defined in (\ref{themonstercorrelationmatrix}).
Since the two matrices ${\bf M}_{\phi,\psi}^{i,j}(k\tau, k\tau')$ defined in (\ref{theMmatrices}) are Hermitian under simultaneous exchange of $i \rightarrow j$ and $t\rightarrow t'$,  we obtain that the two Re$\left[ {\bf M}_{\phi,\psi}^{i,j}(k\tau, k\tau') \right]$ are symmetric, and so are the two ${\bf A}^{i,j}_{\phi,\psi}(x, x')$ operators.

This term gives an imaginary part to the effective action, and therefore cannot be properly interpreted as an ordinary part of the action. Using the Hubbard-Stratonovich transformation \cite{Stratonovich1957,Hubbard:1959ub}, $e^{-ax^2/2}=\sqrt{1/(2\pi a)}\int_{-\infty}^{\infty}e^{-y^2/(2a)-ixy}d y$, we introduce an auxiliary field $y$ obeying Gaussian $a$-statistics and regularly coupling to the initial variable $x$. One can generalize such a formula to symmetric operators and introduce two real classical random fields per field in the system $\xi_1^{\phi}$, $\xi_2^\phi$ and  $\xi_1^{\psi}$, $\xi_2^\psi$, each obeying the Gaussian probability distribution (\ref{noiseprobdistribution}),
and thus rewrite the influence functional as follows (neglecting an overall normalization factor):
\bea
	F[\tilde{\varphi}, \tilde{\chi}]= \exp\left[\frac{i}{\hbar} S^{(1)}_{\uIA} \right] \qquad \qquad \qquad \qquad \qquad \qquad \qquad \qquad \qquad \qquad \qquad \qquad \qquad \qquad \qquad \qquad \qquad \qquad\nonumber \\
	=\int \mathcal{D} \xi^{\phi}_1 \mathcal{D} \xi^{\phi}_2 \exp\left\lbrace\int d^4xd^4x' [\xi^{\phi}_1 (x), \xi^{\phi}_2(x)] {\bf A_\phi}^{-1}(x,x')\left[\begin{array}{c}\xi^{\phi}_1(x')\\  \xi^{\phi}_2(x') \end{array}\right] +\frac{i}{\hbar}\int dx a^3(t)\left[ \sigma^\phi_q(x),\, \varphi_q(x) \right] \left[\begin{array}{c}\xi^{\phi}_1(x)\\  \xi^{\phi}_2(x) \end{array}\right] \right\rbrace\nonumber\\
	\times \int \mathcal{D} \xi^{\psi}_1 \mathcal{D} \xi^{\psi}_2\exp\left[ \phi\leftrightarrow \psi \right] \qquad \qquad \qquad \qquad \qquad \qquad \qquad \qquad \qquad~~ \qquad \qquad \qquad \qquad \qquad \qquad\nonumber\\
	\equiv\int \mathcal{D} \xi^{\phi}_1 \mathcal{D} \xi^{\phi}_2 \mathcal{P}\left[\xi^{\phi}_1, \xi^{\phi}_2 \right] \exp\left\lbrace \frac{i}{\hbar}\int dx a^3(t)\left[ \sigma^\phi_q(x),\, \varphi_q(x) \right] \left[\begin{array}{c}\xi^{\phi}_1(x)\\  \xi^{\phi}_2(x) \end{array}\right] \right\rbrace \times \int \mathcal{D} \xi^{\psi}_1 \mathcal{D} \xi^{\psi}_2\exp\left[ \phi\leftrightarrow \psi \right] \qquad\quad \qquad ~\,\\
	\equiv\left\langle  \exp\left\lbrace \frac{i}{\hbar}\int dx a^3(t)\left[ \sigma^\phi_q(x),\, \varphi_q(x) \right] \right\rbrace \left[\begin{array}{c}\xi^{\phi}_1(x)\\  \xi^{\phi}_2(x) \end{array}\right]  \exp\left\lbrace \frac{i}{\hbar}\int dx a^3(t)\left[ \sigma^\psi_q(x),\, \chi_q(x) \right] \left[\begin{array}{c}\xi^{\psi}_1(x)\\  \xi^{\psi}_2(x) \end{array}\right] \right\rbrace \right\rangle_{\xi^{\phi}_1,\xi^{\phi}_2, \xi^{\psi}_1, \xi^{\psi}_2}\, , \quad\,
\eea
where the two last equations define the probability density $\mathcal{P}\left[\xi^{\phi}_1, \xi^{\phi}_2 \right]$ and the corresponding average $\left\langle\quad\right\rangle_{\xi^{\phi}_1,\xi^{\phi}_2, \xi^{\psi}_1, \xi^{\psi}_2}$. We therefore see how the influence action can be understood as adding an extra contribution averaged over classical noise sources to the total effective action. 

We are now in a position where we can go back to the total effective action. Writing it in terms of the quantum and classical fields instead of the fields on the forward and backward parts of the time contour using (\ref{transfotoclassquant})
\bea
\label{eq:SeffKeld}
 	\int^{\varphi^+}_{\varphi^\pm_i} \mathcal{D}\varphi^\pm\int^{\chi^+}_{\chi^\pm_i}\mathcal{D} \chi^\pm  \exp\left[\frac{i}{\hbar} S^{(1)}_{\ueff}\right]=\int^{\varphi^+}_{\varphi^\pm_i} \mathcal{D}\varphi^\pm\int^{\chi^+}_{\chi^\pm_i}\mathcal{D} \chi^\pm  \exp\left[\frac{i}{\hbar}\left( S_{\usys}^+-S_{\usys}^-+S^{(1)}_{\uIA}\right)\right]\qquad \qquad\qquad\qquad \qquad\qquad\\
 	=\int \mathcal{D}\varphi^{q,c}\mathcal{D} \chi^{q,c} \int\mathcal{D}\xi_1\mathcal{D}\xi_2 \mathcal{P}\left[\xi_1, \xi_2 \right] \exp\left(\frac{i}{\hbar}\int d^4x a^3(t)\left\lbrace\varphi_q\left(\Box-m^2_\Phi\right) \varphi_c +\chi_q\left(\Box-m_\Psi^2\right)\chi_c\right.\right.\nonumber\qquad\qquad\qquad~~~\qquad\\
	\left.\left. -\tilde{V}\left(\varphi_c+\frac{\varphi_q}{2}, \chi_c+\frac{\chi_q}{2}\right) +\tilde{V}\left(\varphi_c-\frac{\varphi_q}{2}, \chi_c-\frac{\chi_q}{2}\right) +\left[ \sigma^\phi_q(x),\, \varphi_q(x) \right] \left[\begin{array}{c}\xi^{\phi}_1(x)\\  \xi^{\phi}_2(x) \end{array}\right] +\left[ \sigma^\psi_q(x),\, \chi_q(x) \right] \left[\begin{array}{c}\xi^\psi_1(x)\\  \xi^\psi_2(x) \end{array}\right] \right\rbrace\right) ,~~\nonumber
\eea
In order to explore the classical limit of the effective action, we proceed to the field rescaling $\varphi_q,\,\chi_q\rightarrow \hbar\varphi_q,\, \hbar\chi_q $. With this redefinition, a first order expansion of the effective action in $\hbar\varphi_q,\, \hbar\chi_q $ corresponds to the $\hbar$-independent, or classical, sector of the action, which allows us to obtain (\ref{classicalfirstorderaction}).

\renewcommand{\theequation}{B \arabic{equation}}
\renewcommand{\thefigure}{B \arabic{figure}}
\setcounter{equation}{0}  
\setcounter{figure}{0}
\setcounter{subsubsection}{0}
\section*{APPENDIX B}  
\Large\begin{center}
{\scshape Perturbative Expansion}
\\
\bigskip
\end{center}
\normalsize

In this appendix, we define the perturbative expansion in terms of the usual sum over loop diagrams of increasing order, which allows us is section IV B-C to perform the path integral over the bath degrees of freedom up to one loop. Using the notation we introduced in (\ref{perturbativeintaction}) - (\ref{lambda2}) of section \ref{subsec:OneLoop}, we can define the free bath action on the forward and backward time contours:
\be
	e^{\frac{i}{\hbar} \left[  S^{0+}_{\ubath}-S_{\ubath}^{0-}\right]}=e^{\frac{i}{\hbar}\int d^4x \frac{1}{2}\left[\tilde\phi_>^T\tilde{\Lambda}{^{(2)}} \tilde\phi_> +\tilde\psi_>^T\tilde{\Lambda}{^{(2)}} \tilde\psi_>  \right]}\, .
\ee

In order to define this perturbative expansion \cite{Calzetta1987, Hu1993}, we introduce four currents, one for each field on the forward part of the time contour, and one per field on the backward part. Using the vector notation $\tilde J_{\phi,\psi}^{T}=[J^+_{\phi,\psi}, ~~-J^-_{\phi,\psi}]$, we obtain the modified, unperturbed influence functional:
\bea
	F^{(1)}[\tilde J_\phi, \tilde J_\psi]&=&\int_{-\infty}^{\infty}d\phi_{>_f}^+ d\psi_{>_f}^+ \int \mathcal{D} \phi_{>}^{\pm}\int\mathcal{D}\psi_{>}^{\pm} \nonumber \\
	&& \quad\times e^{\frac{i}{\hbar} \left[  S^{0+}_{\ubath}-S_{\ubath}^{0-}+\int_{t_i}^{t_f}d^4x\tilde J^T_\phi\tilde\phi_{>}+\int_{t_i}^{t_f}d^4x\tilde J_\psi^T\tilde\psi_{>} \right]}\, .
\label{unperturbedinfluencefunctional}
\eea
Defining the average of an operator depending on the bath degrees of freedom $\hat{Q}[\tilde\phi_>, \tilde\psi_>] =\hat{Q}\left[{\phi}_>^+,{\psi}_>^+, {\phi}_>^-,{\psi}_>^-\right]$:
\bea
\label{eq:defmean}
	\left\langle \hat{Q}[\tilde\phi_>, \tilde\psi_>] \right\rangle_0&=&\int_{-\infty}^{\infty}d\phi_{>_f}^+ d\psi_{>_f}^+ \int \mathcal{D} \phi_{>}^{\pm}\int\mathcal{D}\psi_{>}^{\pm} e^{\frac{i}{\hbar} \left[  \tilde{S}^{0+}_{\ubath}-\tilde{S}_{\ubath}^{0-} \right]}\hat{Q}[\tilde\phi_>, \tilde\psi_>]  \\
	&=&\left.\hat{Q}\left[ \frac{\hbar}{i}\frac{\delta}{\delta J^+_\phi(x)}, \frac{\hbar}{i}\frac{\delta}{\delta J^+_\psi(x)},  -\frac{\hbar}{i}\frac{\delta}{\delta J^-_\phi(x)}, -\frac{\hbar}{i}\frac{\delta}{\delta J^-_\psi(x)}  \right]F^{(1)}[\tilde{J}_\phi, \tilde{J}_\psi]\right|_{\tilde J_\phi= \tilde J_\psi=0}\, ,
\eea
we can write the full influence functional $F$ in terms of $F^{(1)}$:
\bea
F[\tilde\varphi, \tilde \chi] &=& \left\langle  e^{-\frac{i}{\hbar}\left[ \tilde{V}_{\ubath}\left(\phi_{>}^+, \psi_>^+\right) -\tilde{V}_{\ubath}\left(\phi_{>}^-, \psi_>^-\right) +\tilde{V}_{\uint}\left(\varphi^+, \chi^+, \phi_>^+, \psi_>^+\right) -\tilde{V}_{\uint}\left(\varphi^-,\chi^-, \phi_{>}^-, \psi_>^-\right) \right]} \right\rangle_0\\
&=& \exp\left\lbrace -\frac{i}{\hbar}\left[ \tilde{V}_{\ubath}\left(\frac{\hbar}{i}\frac{\delta}{\delta J^+_\phi(s)}, \frac{\hbar}{i}\frac{\delta}{\delta J^+_\psi(s)}\right) -\tilde{V}_{\ubath}\left(-\frac{\hbar}{i}\frac{\delta}{\delta J^-_\phi(s)}, -\frac{\hbar}{i}\frac{\delta}{\delta J^-_\psi(s)}\right) \right.\right. \nonumber \\
	&&\left.\left. +\tilde{V}_{\uint}\left(\varphi^+, \chi^+,\frac{\hbar}{i} \frac{\delta}{\delta J^+_\phi(s)}, \frac{\hbar}{i}\frac{\delta}{\delta J^+_\psi(s)}\right)-\tilde{V}_{\uint}\left(\varphi^-,\chi^-,-\frac{\hbar}{i} \frac{\delta}{\delta J^-_\phi}, -\frac{\hbar}{i}\frac{\delta}{\delta J^-_\psi(x)}\right) \right] \right\rbrace\left.F^{(1)}[\tilde J_\phi, \tilde J_\psi]\right|_{\tilde J_\phi= \tilde J_\psi=0}\, .
\nonumber \eea
Expanding the exponential, we obtain the perturbative expansion we are after:
\bea
\label{perturbativeloopexpansion}
{F[\tilde\varphi, \tilde \chi]} =\qquad\qquad\qquad\qquad\qquad\qquad\qquad\qquad\qquad\qquad\qquad\qquad\qquad\qquad\qquad\qquad\qquad\qquad\qquad\qquad\qquad\qquad\\
	\langle1\rangle_0-\frac{i}{\hbar}\left\{ \left\langle \tilde{V}_{\ubath}(\phi_{>}^+, \psi_>^+) \right\rangle_0  -\left\langle\tilde{V}_{\ubath}(\phi_{>}^-, \psi_>^-)  \right\rangle_0 +\left\langle\tilde{V}_{\uint}(\varphi^+, \chi^+, \phi_>^+, \psi_>^+) \right\rangle_0  
-\left\langle\tilde{V}_{\uint}(\varphi^-,\chi^-, \phi_{>}^-, \psi_>^-)  \right\rangle_0 \right\}\nonumber \qquad
\\  
{-\frac{1}{2\hbar^2}\left\lbrace\left\langle\tilde{V}_{\ubath}^2(\varphi^+, \chi^+, \phi_>^+, \psi_>^+) \right\rangle_0  +\left\langle\tilde{V}_{\ubath}^2(\varphi^-, \chi^-, \phi_>^-, \psi_>^-) \right\rangle_0  
\right.}{\left.
-2\left\langle \tilde{V}_{\ubath}(\varphi^+, \chi^+, \phi_>^+, \psi_>^+)\tilde{V}_{\ubath}(\varphi^-, \chi^-, \phi_>^-, \psi_>^-) \right\rangle_0 \right.}\nonumber
\\  
{\left.+\left\langle\tilde{V}_{\uint}^2(\varphi^+, \chi^+, \phi_>^+, \psi_>^+) \right\rangle_0  +\left\langle\tilde{V}_{\uint}^2(\varphi^-, \chi^-, \phi_>^-, \psi_>^-) \right\rangle_0  
\right.}{\left.
-2\left\langle \tilde{V}_{\uint}(\varphi^+, \chi^+, \phi_>^+, \psi_>^+)\tilde{V}_{\uint}(\varphi^-, \chi^-, \phi_>^-, \psi_>^-) \right\rangle_0 \right\rbrace}\nonumber
+...
\eea
where the ellipses represent terms with at least three powers of $\tilde{V}$
. The terms already written down will give rise to terms similar to the ones we wrote down in (\ref{stochastichybridfulleomphi}) and (\ref{stochastichybridfulleompsi}) and which we argued represented interactions between the quantum fields and the coarse-grained fields (that is, the terms that are on the fourth line of each equation). 


\renewcommand{\theequation}{C \arabic{equation}}
\renewcommand{\thefigure}{C \arabic{figure}}
\setcounter{equation}{0}  
\setcounter{figure}{0}
\setcounter{subsubsection}{0}
\section*{APPENDIX C}  
\Large\begin{center}
{\scshape Perturbative Calculation of the $\ln \Det$ Terms}
\\
\bigskip
\end{center}
\normalsize

Our goal in this appendix will be to expand the determinants appearing in the expression for the free influence functional (\ref{unperturbedinfluencefunctionalTEXT}) in a way that they can be explicitly calculated perturbatively. Doing so we will link this expansion to a diagrammatic expansion, in order to show that including the quadratic terms as part of the free propagators of the bath fields we get the same result as if we had treated the $\frac{g^2}{2}(\varphi^\pm)^2(\psi^\pm)^2$, $\frac{g^2}{2}(\chi^\pm)^2(\phi^\pm)^2$, and $\frac{\lambda}{4} (\chi^\pm)^2(\psi^\pm)^2$ as perturbations and obtained their one-loop contribution to the perturbed influence functional.

In order to evaluate the logarithm of the determinant of the operators defined in (\ref{unperturbedinfluencefunctionalTEXT}), work with the conformal time $\tau$ in coordinates where the Fourier transform of $\Box=\frac{1}{a^2}\left[ -\delta_\tau^2+\nabla^2 \right]$ is the four-norm $k^2/a^2$ In order to simplify the notation of the propagator bellow, we will set $a=1$ for now, but to restore the factors of the scale factor in the calculation performed bellow, one only needs to add a factor of $1/a^2$ every time a four-norm $k^2$ appears. 

We start by rewriting it as a trace, which we can evaluate as the sum of their eigenvalues:
\bea
\label{C1}
		-\frac{1}{2}\mathrm{Tr} \ln (\tilde \Lambda _{\phi}^{(2)})&=&-\frac{\Omega}{2} \int \frac{d^4k }{(2\pi)^4}\ln \left[\left(-k^2 -m_\Phi^2-g^2(\chi^+)^2\right)\left(k^2 +m_\Phi^2+g^2(\chi^-)^2\right)-0\cdot0\right]\, .
\eea
Here the $\Omega$ factor sitting at the front of the four-momenta integral represent the four-dimensional volume over which we are evaluating the path integral. It arises in this form since for simplicity we have written the $\chi$ fields as if they had no time or spatial dependence, which is the case over the {\bf k}-scales we are considering in the momentum integral, but once we re-introduce their spacetime dependence, this factor will transform into a four-dimensional spacetime integral.

Also, we have kept the zero contributions from the off-diagonal terms of $\Lambda^{(0)}_\phi$ because we wish to re-write this expression in terms of bath propagators. Specifically, the Feynman (Dyson) propagators $i\Delta_{\phi, F(D)}$ and Wightman propagators $i\Delta_{\phi, \gtrless}$ satisfy the inhomogeneous and homogeneous Klein-Gordon equations, respectively:
\bea
	\left[\Box_x -m_\Phi^2-g^2(\chi^{+(-)})^2\right]i\Delta_{\phi, F(D)}(x, y)= (-) i\delta^4(x-y)\, , \qquad \left[\Box_x -m_\Phi^2-g^2\left(\chi^{+(-)}\right)^2\right]i\Delta_{\phi, <(>)}(x, y)= 0\, .
\eea
We therefore replace the two `0' in the above equation (\ref{C1}) by the appropriate product of the two Wightman propagators and their kernels, recalling that they come form the explicit calculation of the determinant. We want to expand this expression around the constant free, massive Klein-Gordon propagator (free of system fields dependence), therefore we factor out those constant leading factors and collect them as follow:
\bea
		-\frac{1}{2}\mathrm{Tr} \ln (\tilde \Lambda _{\phi}^{(2)})&=&-\frac{\Omega}{2} \int \frac{d^4k }{(2\pi)^4} \ln \left\lbrace\left(-k^2 -m_\Phi^2\right)\left(k^2 +m_\Phi^2\right)\left [1 +\frac{-1}{-k^2 -m_\Phi^2}g^2(\chi^+)^2\right]\left[1+\frac{1}{k^2 +m_\Phi^2}g^2(\chi^-)^2\right]  \right. \nonumber\\
		&&\left. - (-k^2 -m_\Phi^2)(k^2 +m_\Phi^2)\left[1+\frac{-1}{-k^2 -m_\Phi^2}g^2\left(\chi^{+}\right)^2\right]i\Delta_{\phi, <}(k)\left[1+\frac{1}{k^2 +m_\Phi^2}g^2\left(\chi^{-}\right)^2\right]i\Delta_{\phi, >}(k) \right\rbrace\nonumber\, .\\
\eea

Looking at the Wightman propagators appearing in the above, we would like to write them as an expansion around the free, massive, system fields-independent ones, $\Delta^{(0)}_{\phi, \gtrless}(k)$. Since $\Delta_{\phi, \gtrless}(k)$ are each proportional to a delta function in Fourier space, and therefore their Fourier expansion should be interpreted as a requirement to integrate by parts the Fourier integral appearing at the front of the above equation, once we perform that expansion we obtain $\Delta_{\phi, \gtrless}(k)= \Delta^{(0)}_{\phi, \gtrless}(k)+\mathcal{O}(g^4)$ plus other terms which will give zero as a $c$-number once we have imposed the fact that free Wightman propagators are solutions to the free homogeneous Klein-Gordon equation: $(k^2+m_\Phi^2)i\Delta^{(0)}_{\phi, \gtrless}(k)=0$. Simplifying, we obtain:

\bea
		-\frac{1}{2}\mathrm{Tr} \ln (\tilde \Lambda _{\phi}^{(2)})&=&-\frac{\Omega}{2} \int \frac{d^4k }{(2\pi)^4}\ln \left[\left(-k^2 -m_\Phi^2\right)\left(k^2 +m_\Phi^2\right)\right]  + \nonumber \\
		&& \ln \left [1 +\frac{-g^2(\chi^+)^2}{(-k^2 -m_\Phi^2)}+\frac{g^2(\chi^-)^2}{(k^2 +m_\Phi^2)}+\frac{-g^2(\chi^+)^2}{(-k^2 -m_\Phi^2)}\frac{g^2(\chi^-)^2}{(k^2 +m_\Phi^2)}- \frac{0\cdot (-g^2(\chi^{+})^2)}{(-k^2-m_\Phi^2)}\cdot\frac{0\cdot( g^2(\chi^-)^2)}{(k^2+m_\Phi^2)}\right]\nonumber\\
		&=&-\frac{1}{2}\mathrm{Tr} \ln (\tilde \Lambda _{\phi}^{(0)})+\frac{\Omega}{2} \int \frac{d^4k }{(2\pi)^4} \left\lbrace\frac{g^2(\chi^+)^2}{(-k^2 -m_\Phi^2)}-\frac{g^2(\chi^-)^2 }{(k^2 +m_\Phi^2)}- \frac{0\cdot (g^2(\chi^{+})^2)}{(-k^2-m_\Phi^2)}\cdot\frac{0\cdot(g^2(\chi^-)^2)}{(k^2+m_\Phi^2)}\right.\nonumber \\
		&&\qquad\left. +\frac{1}{2}\left[\frac{g^2(\chi^+)^2}{(-k^2 -m_\Phi^2)}\cdot\frac{g^2(\chi^+)^2}{(-k^2 -m_\Phi^2)} +\frac{g^2(\chi^-)^2}{(k^2 +m_\Phi^2)}\cdot \frac{g^2(\chi^-)^2}{(k^2 +m_\Phi^2)}\right]+...\right\rbrace\, ,\nonumber
\eea
where in the last line we have used our assumption of the validity of perturbation theory around small coupling constants to Taylor expand the $\ln$'s up to $\mathcal{O}(g^2)$ and where $ \tilde\Lambda _{\phi}^{(0)}$ [as defined in (\ref{matrixnotationshort}) and (\ref{integrationkernelsforscalarfields})] is the integration kernel for a free, massive scalar field and so has no system field dependence, which means the terms we have collected in a new $\mathrm{Tr} \ln $ term boil down to being simply an overall constant multiplying the influence functional. 

Next we want to re-write this expression by introducing factors of the free bath propagators. To do so, we use the fact that free propagators are solutions to the Klein-Gordon equation:
\bea
	(\Box_x-m_\Phi^2)\left[ \begin{array}{c} i\Delta^{(0)}_{\phi,F} \\ i\Delta^{(0)}_{\phi,D}\\ -i\Delta^{(0)}_{\phi,<} \\ -i\Delta^{(0)}_{\phi,>} \end{array} \right]=\left[ \begin{array}{c} i\\ -i \\0 \\0 \end{array}\right] \delta^4(x-x')\, .
\eea

We obtain:
\bea
		-\frac{1}{2}\mathrm{Tr} \ln (\tilde \Lambda _{\phi}^{(2)})=-\frac{1}{2}\mathrm{Tr} \ln (\tilde \Lambda _{\phi}^{(0)})+\frac{\Omega}{2} \int \frac{d^4k }{(2\pi)^4} \left[g^2\left(\Delta^{(0)}_{\phi,F}(k) (\chi^+)^2+\Delta^{(0)}_{\phi,D}(k) (\chi^-)^2\right)\right.\qquad\qquad\qquad\qquad\qquad\qquad\qquad\\
		\left. +\frac{g^4}{2}\left[\Delta^{(0)}_{\phi,F}(k)(\chi^+)^2\Delta^{(0)}_{\phi,F}(k) (\chi^+)^2  - 2\Delta^{(0)}_{\phi,<}(k) (\chi^{+})^2 \Delta^{(0)}_{\phi,>}(k)\rangle (\chi^-)^2+\Delta^{(0)}_{\phi,D}(k) (\chi^-)^2\Delta^{(0)}_{\phi,D}(k) (\chi^-)^2\right]+...\right]\, .\nonumber
\eea
As advertised at the beginning of the appendix, this result matches exactly the contributions coming from one-loop diagrams up to $\mathcal{O}(g^4)$ (which is equivalent to $\mathcal{O}\left[(\phi^\pm)^2\right]$ in terms of the bath degrees of freedom) we would have obtained if we had treated the $\frac{g^2}{2}(\phi^\pm)^2(\chi^\pm)^2$ term in the action perturbatively instead of as a time-dependent mass for the bath fields.
Indeed, the first two terms in the above expression represent contributions from the following two diagrams:
\begin{figure}[h]
\begin{center}
\includegraphics[width=5.5cm]{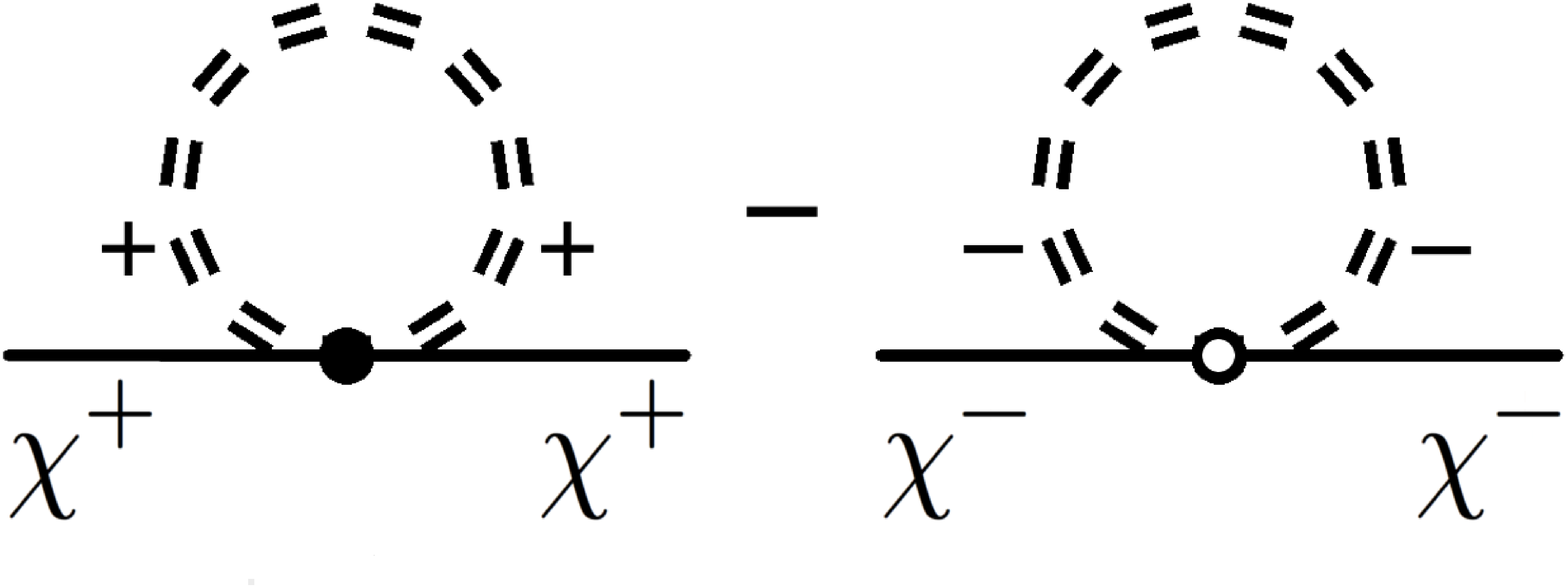}
\label{fig:oneloop}
\end{center}
\end{figure}
whereas the remaining terms are the contributions coming from the following $\mathcal{O}(g^4)$ (which is equivalent to $\mathcal{O}\left[(\phi^\pm)^4\right]$ in the bath degrees of freedom) diagrams:
\begin{figure}[h]
\begin{center}
\includegraphics[width=15cm]{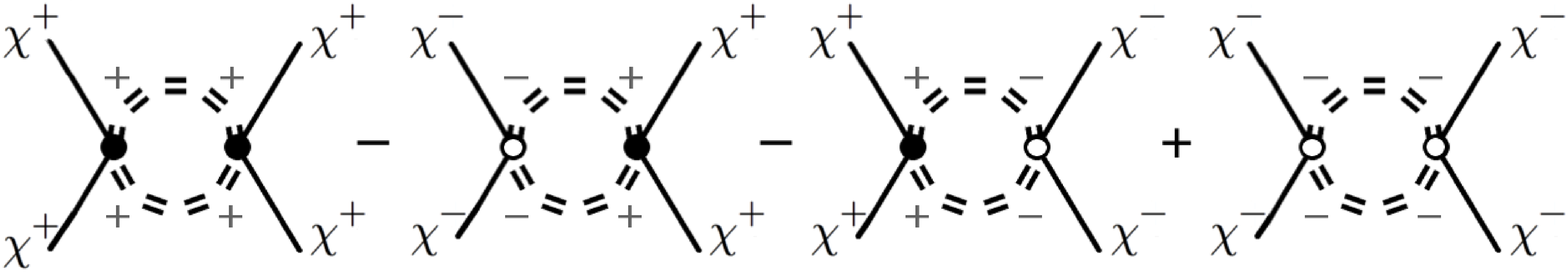}
\label{fig:oneloop}
\end{center}
\end{figure}

Reintroducing the spacetime dependence of the system fields, and rewriting the propagators in position space and in matrix form, we obtain:
\bea
	-\frac{1}{2}\mathrm{Tr} \ln (\tilde \Lambda _{\phi}^{(2)})=-\frac{1}{2}\mathrm{Tr} \ln (\tilde \Lambda _{\phi}^{(0)})-\frac{i}{\hbar}\frac{g^2 }{2} \int d^4x  \left[ \begin{array}{cc} \chi^+(x),& \chi^-(x) \end{array}\right] \left( \begin{array}{cc}T \langle \hat{\phi}^+_x\hat{\phi}^+_x \rangle&0 \\0&\bar{T}\langle \hat{\phi}^-_x\hat{\phi}^-_x\rangle \end{array}\right)\left[ \begin{array}{c} \chi^+(x) \\ \chi^-(x) \end{array}\right]\qquad\qquad\\
		 +\left(\frac{i}{\hbar}\right)^2\frac{g^4}{4} \int d^4 xd^4x'   \left[ \begin{array}{cc}  (\chi^{+}(x))^2,& (\chi^{-}(x))^2 \end{array}\right]  \left[ \begin{array}{cc}\left( T  \langle \hat{\phi}^+_x\hat{\phi}^+_{x'} \rangle\right)^2 &   - \left( \langle \hat{\phi}^-_{x'}\phi^+_x \rangle\right)^2   \\ - \left( \langle \hat{\phi}^-_{x}\phi^+_{x'} \rangle\right)^2  & \left(\bar{T} \langle \hat{\phi}^-_x\hat{\phi}^-_{x'}\rangle \right)^2 \end{array}\right]\left[ \begin{array}{c} (\chi^+(x'))^2 \\ (\chi^-(x))^2 \end{array}\right]+...\, .\nonumber
\eea
Rotating to the Keldysh basis, we finally obtain (\ref{detlogpsi}).

To derive (\ref{detlogphi}), one proceeds in an analogous manner, however one obtains more contributions at every order in the couplings, since there are two time-dependent mass terms in the non-zero entries of $\Lambda^{(2)}_\psi$ instead of one. The additional diagrammatic contributions we obtain are, at $\mathcal{O}(g^2)$ or $\mathcal{O}(\lambda)$ in the couplings (i.e. $\mathcal{O}\left[(\psi^\pm)^2\right]$ or $\mathcal{O}\left[(\phi^\pm)^2\right]$ in the bath degrees of freedom): 
%

%
\begin{figure}[h]
\begin{center}
\includegraphics[width=6.5cm]{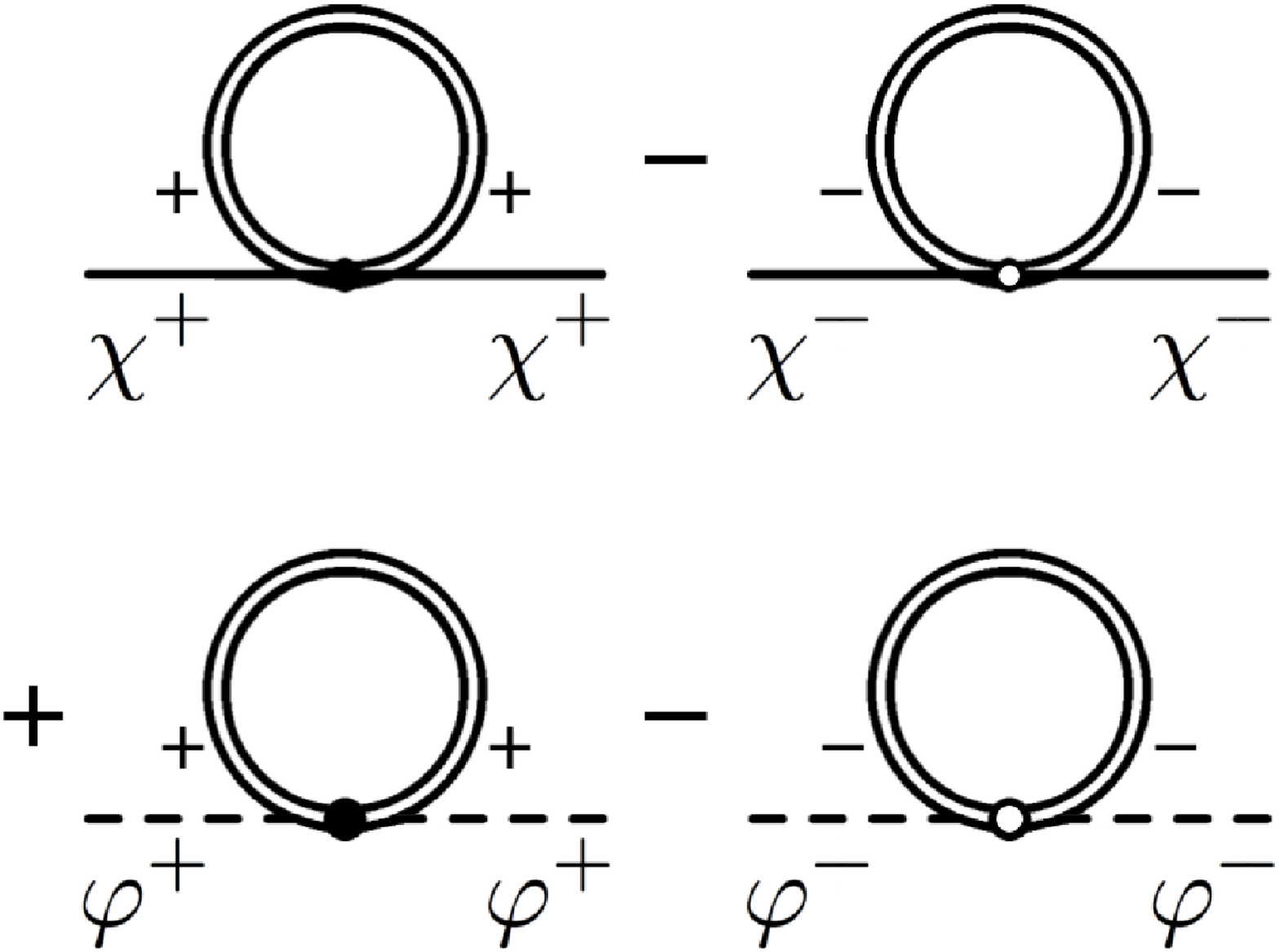}
\label{fig:oneloop}
\end{center}
\end{figure}

while the terms coming from the expansion of the $\ln \Det(\Lambda^{(2)}_\psi)^{-1}$ term giving contributions at $\mathcal{O}(g^4)$ or $\mathcal{O}(\lambda^2)$ (or equivalently $\mathcal{O}\left[(\psi^\pm)^4\right]$ or $\mathcal{O}\left[(\phi^\pm)^4\right]$) are equivalent to the contribution coming form the following diagrams:
\\

\begin{figure}[h]
\begin{center}
\includegraphics[width=15cm]{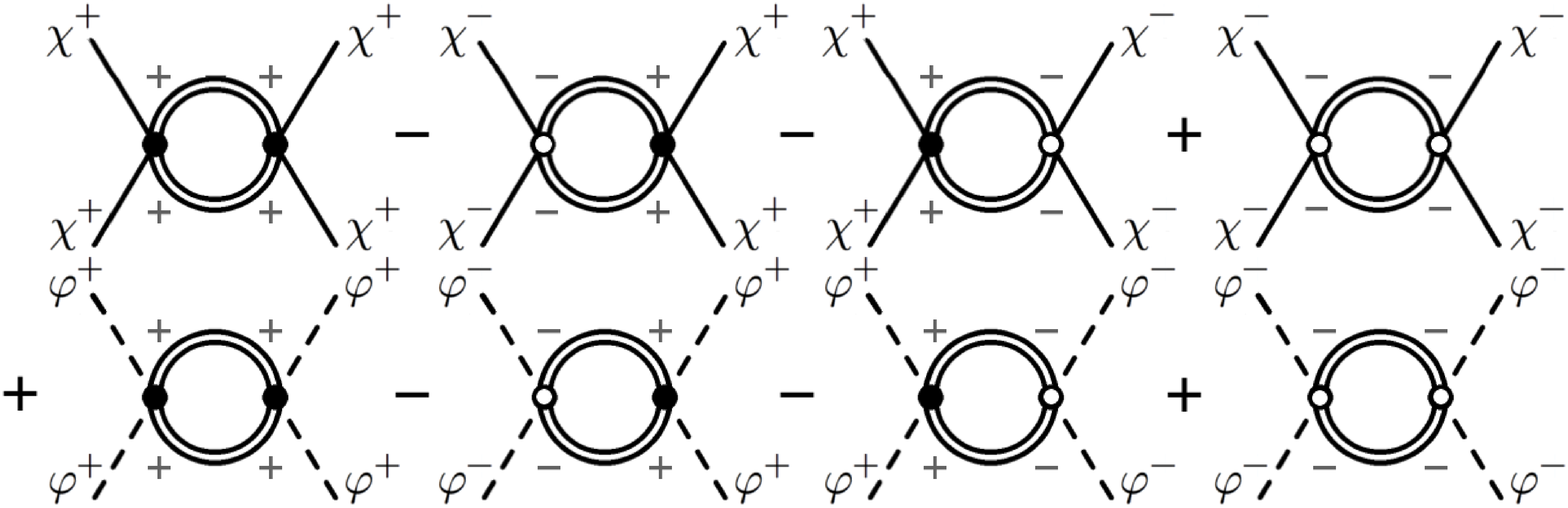}
\label{fig:oneloop}
\end{center}
\end{figure}

\newpage
\end{widetext}

\bibliography{biblio2}

\end{document}